\documentclass[aps,prd,preprint,tightening,showpacs]{revtex4-1}
\usepackage{amsmath,amssymb,amsthm,graphicx}
\usepackage[mathscr]{eucal}
\usepackage[colorlinks,linkcolor=blue,anchorcolor=blue,citecolor=green]{hyperref}
\usepackage[dvipsnames,usenames]{color}
\usepackage{graphicx}
\usepackage{subfigure}
\usepackage{amsmath}

\newcommand{\be}{\begin{equation}}
\newcommand{\ee}{\end{equation}}
\newcommand{\ba}{\begin{eqnarray}}
\newcommand{\ea}{\end{eqnarray}}
\usepackage{epsf,epsfig,graphics}
\usepackage{verbatim,color,ulem}
\usepackage{physics}
\begin{document}
\title{Throat effects on strong gravitational lensing in Kerr-like wormholes}
\author{Tien Hsieh}
\author{Da-Shin Lee}
\email{dslee@gms.ndhu.edu.tw}
\author{Chi-Yong Lin}
\email{lcyong@gms.ndhu.edu.tw}
\affiliation{
Department of Physics, National Dong Hwa University, Hualien 97401, Taiwan, Republic of China}
\date{\today}

\begin{abstract}
We study strong gravitational lensing by a specific one-parameter extension of Kerr spacetime, a Kerr-like wormhole, characterized by a single parameter specifying the throat's location.
We classify the roots of the radial potential derived from the null geodesic equations.
We focus on the conditions required for the throat, together with the other roots, to become either a double root or a triple root, potentially leading to the divergence of the deflection angle of the light rays in the strong deflection limit (SDL).
In particular, while a logarithmic divergence of the deflection angle is known to occur as the closest distance $r_0$ of an incident light trajectory around a black hole approaches a double root, a stronger power-law (nonlogarithmic) divergence is found as $r_0$ approaches a triple root especially in a wormhole.
In addition, the effective potential in terms of the proper distance from the throat is constructed, with which one can see how the light rays can either travel within a single spacetime, where both the source and the observers are located, or pass from the source through the throat into another spacetime where different observers reside.
%
Observational effects, such as relativistic images resulting from the deflection of light by wormholes, are discussed, and they could serve as a unique feature of wormholes.
\end{abstract}


\maketitle

\section{Introduction}
Wormholes are a hypothetical spacetime connecting two separate regions of the Universe or even different universes.
These hypothetical structures in spacetime are solutions to Einstein's equations of general relativity \cite{misner-1974}.
Einstein and Rosen introduced a mathematical construction in order to eliminate coordinate or curvature singularities with a bridgelike structure known as the Einstein-Rosen bridges \cite{einstein-1935}.
Wheeler later interpreted the Einstein-Rosen bridge as a link between the distant points in spacetime, and then coined the term wormhole, albeit on a small scale \cite{wheeler-1955, wheeler-1962}.
Traversable wormholes were proposed by Morris and Thorne \cite{morris-1988}, allowing the observers on a human scale to freely traverse them.
Since then, numerous publications have explored all sorts of wormholes, all of which have one thing in common: they violate the energy condition, at least in a neighborhood of the wormhole throat.
For a review, see \cite{Visser-1995}.
While the exotic stabilization schemes have been proposed, there is currently no evidence that wormholes exist.

Einstein's century-old prediction as a consequence of general relativity (GR) has been confirmed by the recent detection of gravitational waves emitted by the merger of the binary black holes \cite{abbott-2016, abbott-2019, abbott-2021}.
The capture of the spectacular images of a supermassive black hole M87* at the center of the M87 galaxy \cite{akiyama-2019}
and Sgr A* at the center of our galaxy \cite{collaboration-2022}
provide further an important scientific evidence directly confirming the existence of black holes.
Gravitational lensing is the other powerful tool for testing GR \cite{misner-1974, hartle-2003} especially from strong field perspectives \cite{virbhadra-2000, frittelli-2000, bozza-2001, bozza-2002, eiroa-2002, iyer-2007, iyer-2009, tsukamoto-2017, hsiao-2020, hsieh-2021A, hsieh-2021B}.
In \cite{hsiao-2020, hsieh-2021A, hsieh-2021B}, we have studied the gravitational lensing and time delay by a spinning charged black hole.
About the wormholes, Damour and Solodukhin were the first to propose the Schwarzschild-like wormhole \cite{damour-2007}, where the gravitational lensing was studied in the strong deflection limit (SDL) \cite{tsukamoto-2020, ovgun-2018}.
Following the same construction, Kerr-like wormholes were also proposed.
Their throat effects have been extensively explored in relation to the shadow image and the echo \cite{kasuya-2021, bueno-2018, amir-2019}.
Here we will focus on gravitational lensing by the Kerr-like wormholes.
The observational consequences will be compared with the Kerr black holes in \cite{hsieh-2021A}, so that we might be able to tell the difference between wormholes and black holes.
Notice that we consider a particular type of wormhole with one additional parameter $\lambda$ compared to the standard Kerr black hole \cite{carter-1968, frolov-2007, frolov-2017, visser-2023}.
Reduction of the more general extension of the Kerr spacetime to the specific one-parameter wormhole model in this paper will be discussed in Appendix \ref{secV_add}.

\begin{figure}[htp]
\begin{center}
    \includegraphics[width=15cm]{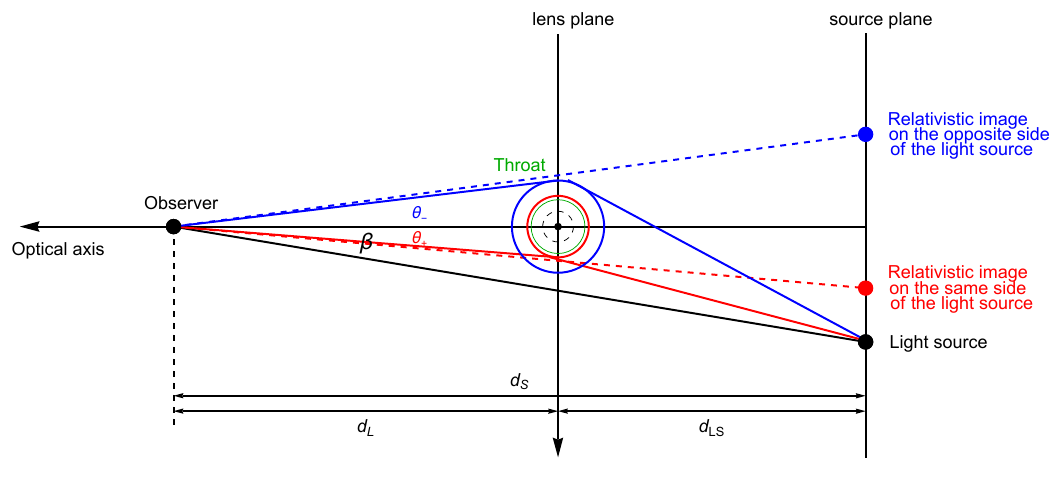}
    \caption{
    Gravitational lens and relativistic images when the observers and the light sources are in the same spacetime.
    Considering the Kerr-like wormhole with angular momentum of the clockwise rotation, the light ray is emitted from the source, and circles around the wormhole multiple times in the SDL along a direct orbit (red line) or a retrograde orbit (blue line). The graph illustrates two sets of the relativistic images.
    When the observer and the light source are in the different spacetimes, mapping the light ray in the spacetime of the source to that of the observer in terms of the proper distance measured from the throat $l$ through $l \to -l$, will expect to see the same lens diagram above ( for example, see Fig. 6 of \cite{taylor-2014}). See also the main text for details.}
    \label{fig:arch_02}
\end{center}
\end{figure}

The lensing effects due to the wormholes can be examined as illustrated in Fig. \ref{fig:arch_02}.
The light rays are emitted from the source and circle around the wormholes multiple times in the SDL along a direct orbit (red line) or a retrograde orbit (blue line), giving two sets of relativistic images.
However,
a unique feature of wormholes is that, for certain parameters of the light rays, they can pass through the throat into another spacetime to be observed.
%
%
%
Due to the symmetry of the proper distance $l$ measured from the throat, which will be defined later, i.e., $l\to -l$, one can map the light rays from the spacetime of the sources to another spacetime of the observers.
This type of trajectories of the light traveling between two spacetimes can be visualized in Fig. 6 in \cite{taylor-2014}.
Thus, according to \cite{shaikh-2019} the lens equation, where the observers and the light sources are in the same spacetime, is assumed to apply to the case, where the observers and the light sources are in different spacetimes.
In the future work, we will extend the work of \cite{taylor-2014} to find the detailed trajectories of the light rays and to further depict the light deflection by the wormholes in a single spacetime or between two spacetimes given in what we call the embedded diagram in Fig. \ref{embedded_diagram}.
Following the approach of \cite{hsieh-2021A} enables us to look for the observational consequences due to the gravitational lensing by wormholes.
Our findings will be useful to constrain the parameter $\lambda$ by the observations at the horizon scales \cite{tsukamoto-2023, vagnozzi-2023}.
In the near future, astrophysical observations will be sufficiently developed to allow very sensitive searches for wormholes in the astrophysical environment \cite{dai-2019, simonetti-2021, bambi-2021}.
\\
%

Layout of the paper is as follows.
In Sec. \ref{secII}, we first introduce the Kerr-like wormholes.
Then, we give a brief review of the equations of motion for the light, mainly focusing on the equatorial orbit, from which the radial potential can be defined in terms of its roots.
Thus, the effective potentials, as a function of the coordinate $r$ and the proper distance $l$, respectively, are constructed to clearly reveal various types of light deflection under investigation.
Section \ref{secIII} is devoted to the study of the strong field limit, where the throat of the wormhole together with other roots of the radial potential become a double or a triple root, giving the divergence of the deflection angle as the root is approached by the closest distance of the incident light ray around a wormhole.
In Sec. \ref{secIV}, the deflection angle in the weak field limit is also studied when the turning point is far from the throat.
The relativistic images of gravitational lens, leading to observational effects, are shown in Sec. \ref{secV}.
In Sec. \ref{secVI}, we summarize the main results and discuss the perspectives.
Appendix \ref{app} is added to summarize the roots of the radial potential of Kerr black holes, which are also the roots of the radial potential of Kerr-like wormholes.
Reduction to the metric we consider from the more general Kerr spacetime is discussed in Appendix \ref{secV_add}.

\section{Kerr-like wormholes and the deflection angle}
\label{secII}
\begin{figure}[htp]
\begin{center}
    \includegraphics[width=10cm]{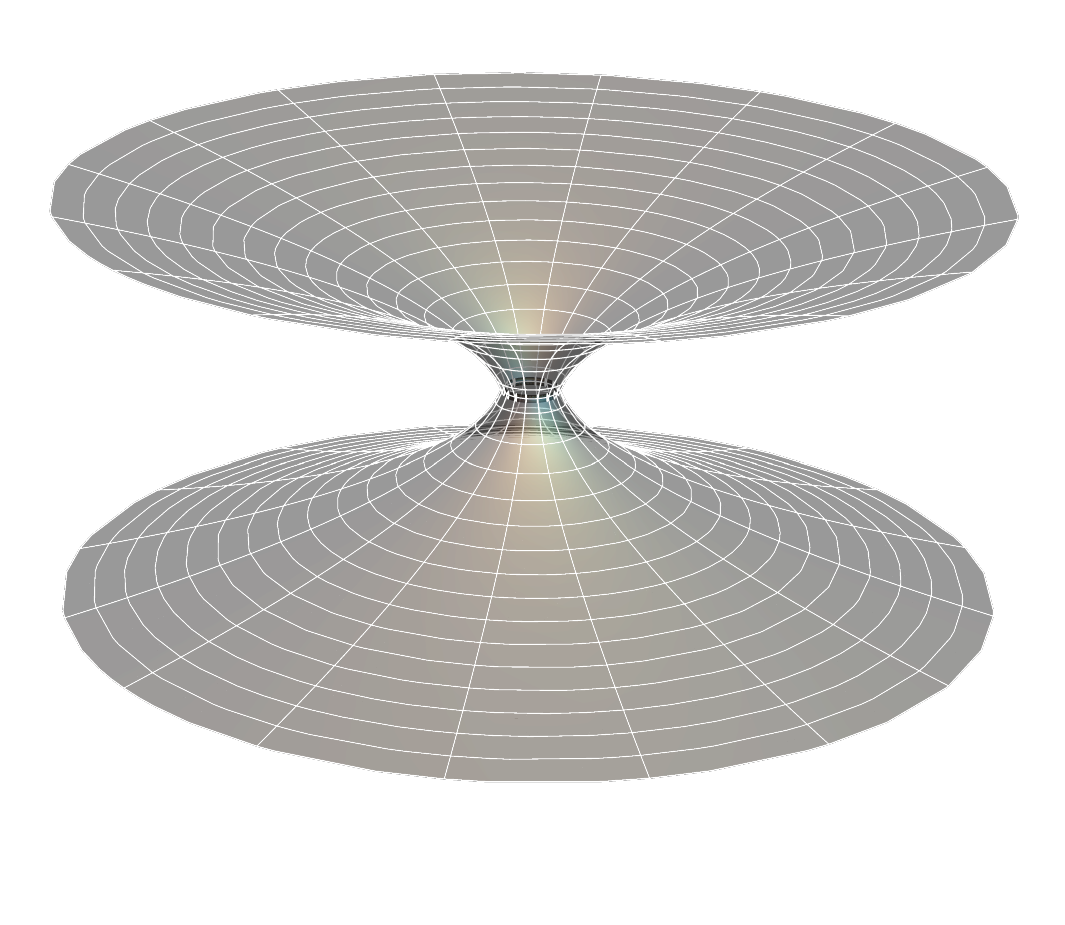}
    \caption{
    Embedding diagram of the Kerr-like wormhole with $a/M=0.5$ and $\lambda=0.3$ \cite{amir-2019}. The upper spacetime corresponds to the proper radial distance $l>0$ while the lower one corresponds to $l<0$. The thinnest place between two spacetimes is the throat at $l=0$ in (\ref{proper_l}).}
    \label{embedded_diagram}
\end{center}
\end{figure}

We first introduce the Kerr-like wormhole metric with which to consider the deflection angle $\hat{\alpha}(b)$ of light rays for a given impact parameter $b$, in particular in the SDL.
The metric of the Kerr-like wormhole is constructed by an additional parameter $\lambda$ characterizing a throat at $r_{\rm th}$ from that of the Kerr black hole with a family of the parameters of the gravitational mass $M$ and angular momentum per unit mass $ a=J/M$, which reads, in Boyer-Lindquist coordinates, as \cite{bueno-2018}
\ba \label{KN_metric}
{ds}^2 &=& g_{\mu\nu} dx^\mu dx^\nu \nonumber \\
&=& -\left( 1-\frac{2Mr}{\Sigma} \right){dt}^2 -\frac{ 2M a r \sin ^2\theta }{\Sigma } ({dt}{d\phi+ d\phi dt)} \nonumber\\
&&
+\frac{\Sigma}{\hat{\Delta}} dr^2 +\Sigma{\, d\theta}^2 + \left( r^2+a^2 + \frac{2Ma^2r \sin^2\theta}{\Sigma} \right)\sin^2\theta {d\phi}^2 \, ,
\ea
where
\be \label{Delta}
\begin{split}
\Sigma &=r^2+a^2\cos^2\theta, \\
\hat{\Delta} &=r^2+a^2-2 M (1+\lambda^2) r =(r-r_{\rm th})(r-\tilde{r}_-), \\
\end{split}
\ee
and the parameter $\lambda \geq 0$ in general.
The radius of the throat can be obtained by solving $\hat{\Delta}(r)=0$, giving
\be \label{r_throat}
r_{\rm th} =M(1+\lambda^2)+\sqrt{M^2(1+\lambda^2)^2 -a^2},
\ee
and another root is
\be \label{r_root}
\tilde{r}_- =M(1+\lambda^2)-\sqrt{M^2(1+\lambda^2)^2 -a^2}.
\ee
The Schwarzschild-like wormhole \cite{damour-2007} can be reduced by setting $a = 0$.
However, in the case of $\lambda=0$, $\hat{\Delta}$ reduces to $ \Delta$ of the Kerr black hole, and becomes
\be
\Delta =r^2+a^2-2 M r =(r-r_+)(r-r_-) \\
\ee
with the outer (inner) horizon $r_+$ ($r_-$) given by
\be \label{rpm_k}
r_{\pm} = M\pm\sqrt{M^2-a^2}\,
\ee
in the condition of $M^2 > a^2$.
Additionally, the ADM mass of the wormhole can be computed to be \cite{amir-2019}
\begin{equation}
M^{(\text{ADM})}_{\text{WH}} = M (1 + \lambda^2)
\end{equation}
as seen by an observer at asymptotic spatial infinity.

For this asymptotically flat, stationary, and axial-symmetric spacetime, where the metric is independent of $t$ and $\phi$, there exist the conserved quantities, namely the energy $\varepsilon$ and the azimuthal angular momentum $\ell$ along a geodesic.
These conserved quantities can be constructed by the Killing vectors and the 4-velocity $u^\mu = {dx^\mu}/{d\sigma} \equiv \dot{x}^\mu$ in terms of the affine parameter $\sigma$ as
\begin{align}
\varepsilon &\equiv -\xi^\mu_t u_\mu, \quad \ell \equiv \xi^\mu_\phi u_\mu \,
\end{align}
with the associated Killing vectors given by
\begin{align}
\begin{aligned}
\xi^\mu_t &= \delta^\mu_t, \quad \xi^\mu_\phi = \delta^\mu_\phi \, .
\end{aligned}
\end{align}
Here, we focus on the light rays traveling on the equatorial plane of the wormhole for $\theta=\pi/2$ and $\dot{\theta}=0$.
Thus, together with the null geodesics, where $u^\mu u_\mu =0$, the equations of motion in the Boyer-Lindquist coordinates can be written as the first-order differential equations by following \cite{gralla-2018, gralla-2020, wang-2022}.
The equations for $\phi (\sigma)$ and $t(\sigma)$ are the same as those in the Kerr black hole given by \cite{gralla-2018, gralla-2020, wang-2022}
\begin{align}
&\frac{\Sigma}{\varepsilon}\frac{d\phi}{d\sigma}=\frac{a}{\Delta}\left(r^2+a^2-ab_s\right)
+{b_s}-a \, ,\label{phi_eq}\\
&\frac{\Sigma}{\varepsilon}\frac{dt}{d\sigma}=\frac{r^2+a^2}{\Delta}\left(r^2+a^2-a b_s\right)+
a\left(b_s-a\right) \, \label{t_eq}
\end{align}
with $\Sigma =r^2$ on the equatorial plane.
In these equations, we have introduced the impact parameter
\begin{align}
b_s=s\left|\frac{\ell}{\varepsilon}\right|\equiv s\,b \;
\end{align}
with $s=\text{Sign$(\ell/\varepsilon)$}$ and $b$ being a positive magnitude.
The parameter $s=+1$ for $b_s>0$ is referred to as a direct orbit, and those with $s=-1$ for $b_s<0$ as a retrograde orbit (see Fig. \ref{fig:arch_02} for the sign convention).
The equation of motion along the $r$ direction is given by
\be
\frac{\Sigma}{\varepsilon}\frac{dr}{d\sigma}=\pm_r\sqrt{{\tilde R}(r)}\, , \label{r_eq}\\
\ee
where the radial potential of a Kerr-like wormhole $\tilde{R}(r)$ with the $\lambda$ dependence is obtained as \cite{amir-2019, kasuya-2021}
\be \label{R_tilde}
\tilde{R}(r) = \frac{\hat{\Delta}(r)}{\Delta(r)} \left[ r^4 + (a^2-b^2) r^2 + 2M(b_s-a)^2 r \right]
= \frac{\hat{\Delta}(r)}{\Delta(r)} R(r) \,.
\ee
One can also write $\tilde{R}(r)$ in terms of its counterpart $R(r)$ for the Kerr black hole, expressed in Appendix \ref{app}, which depends solely on the mass $M$ and the spin $a$. The radial potential $R(r)$ has four roots, explicitly written as $R(r) =(r-r_1)(r-r_2)(r-r_3)(r-r_4)$ in \cite{wang-2022}, where $r_1 < r_2 < r_3 \le r_4$ when the roots are all real valued.
Notice that $r_2=0$ for the equatorial orbits.

To find the closest distance of the light rays $r_0$, we start from solving the radial potential $\tilde{R}(r_0)=0$, leading to two possible solutions, $r_0=r_{\rm th}$ or $R(r_0)=0$.
The former means that light rays will reach the throat, fall into the wormhole, and then travel to another spacetime; the latter is that the outermost root $r_4$ is a turning point $r_0$ of the incident light rays with the value given by
\be \label{r0b}
r_0(b) = 2 \sqrt{\frac{1}{3} \left(b^2-a^2\right)} \cos \left\{ \frac{1}{3} \cos^{-1}\left[ \frac{ -a^2 M +2M sa b -b^2 M}{\left( \frac{b^2-a^2}{3}\right)^{3/2}} \right] \right\}\, ,
\ee
which can be obtained from $R(r_0)=0$ or (\ref{r_4}) in Appendix by rewriting it in a form of trigonometric functions.
One also can obtain the corresponding impact parameter as
\be \label{br0_k}
b(r_0) =\frac{2sMa -r_0 \sqrt{a^2-2r_0 M+r_0^2}}{2M-r_0}\,.
\ee
%
The equation of motion along the radial direction in (\ref{r_eq}) can be cast in the form \cite{hsiao-2020}
\be \label{V_eff}
\frac{1}{b^2} =\frac{\dot{r}^2}{\ell^2} +W_\text{eff}(r)\; ,
\ee
from which to define the effective potential $W_\text{eff}$ as
\be
W_\text{eff}(r) = \left[ 1- \frac{(r-r_{\rm th})(r-\tilde{r}_-) }{(r-r_+)(r-r_-)} \right] \frac{1}{b^2} + \frac{(r-r_{\rm th})(r-\tilde{r}_-) }{(r-r_+)(r-r_-)} \frac{1}{r^2}\left[1-\frac{a^2}{b^2}-\frac{2M }{r}\left(1-\frac{s a}{b}\right)^2\right] \, ,
\ee
with the graphs in Fig. \ref{effW}.
The effective potential diagram clearly helps us to understand the various types of orbits of light traveling outside the throat of a wormhole.
For more details about Fig. \ref{effW}, the red curve represents the effective potential where light can come from spatial infinity, meet the turning point $r_0 = r_4$, and then return to spatial infinity.
The effective potential in a blue curve shows the existence of unstable circular motion with a radius of $r_4 = r_3 = r_{sc}$. The green (and purple) curves depict the effective potential where incident light rays can approach the throat of wormhole and eventually fall into the throat.
In contrast to $W_\text{eff}(r)$ as a function of the coordinate $r$ within a single spacetime, in order to study the light rays traveling between two spacetimes, an alternative coordinate, called the proper radial distance $l$, is often introduced, giving an effective potential denoted by $w_{\text{eff}}(l)$.
We now introduce the proper radial distance $l$ connecting two spacetimes by
\begin{eqnarray}
l &&\equiv \pm \int_{r_{\rm th}}^r \sqrt{\frac{\Sigma}{\hat{\Delta}}} dr \nonumber\\
&&=\sqrt{r^2+a^2-2M(1+\lambda^2)r} +M(1+\lambda^2)
\log{\left( \frac{\sqrt{M^2(1+\lambda^2)^2-a^2}}{r-M(1+\lambda^2)-\sqrt{r^2+a^2-2M(1+\lambda^2)r}} \right)},\nonumber\\
\label{proper_l}
\end{eqnarray}
where the value of $l$ can be extended to $-\infty < l < \infty$, and the throat is located at $l=0$.
In Fig. \ref{embedded_diagram}, the embedded diagram of a Kerr-like wormhole on the equatorial plane is drawn \cite{amir-2019}, where $l >0$ corresponds to one spacetime and $l<0$ another spacetime.
Then, Eq. (\ref{V_eff}) can be rewritten in terms of the proper radial distance $l$ as
\be \label{w_eff}
\dot{l}^2 + \varepsilon^2 w_\text{eff}(l) =0
\ee
with $w_\text{eff}$ given by
\be
w_\text{eff}(l)= \frac{ 4 M sa b +b^2 [r(l)-2 M] -a^2 [2 M+r(l)] -r^3(l)}{r(l) \left[a^2+ r^2(l) -2 M r(l)\right]} \, .
\ee
The graphs of $w_\text{eff}(l)$ are shown in Fig. \ref{effV}.
In particular, the orbits of light rays with the effective potential $W_\text{eff}(r)$ of the green and purple curves in Fig. \ref{effW} can be further understood by the effective potential $w_\text{eff}(l)$ in Fig. \ref{effV} also of the green and purple curves that the light reaches the throat at $l = 0$ and subsequently enters another spacetime.

Finally, solving the differential equations of (\ref{r_eq}) and (\ref{phi_eq})
\be\label{dr/dphi}
\frac{dr}{d\phi} = \frac{\sqrt{(r-r_+)(r-r_-)(r-r_{\rm th})(r-\tilde{r}_-) R(r)}}{br(r-2M)+2Msar}
\ee
gives the deflection angle $\hat{\alpha}$,
\be \label{deflection_angle}
\hat{\alpha}+\pi = 2 \int_{r_0}^{\infty} \frac{br(r-2M)+2Mar}{\sqrt{(r-r_+)(r-r_-)(r-r_{\rm th})(r-\tilde{r}_-) R(r)}} dr.
\ee
In the next section, we will calculate the deflection angle with this formula in the SDL.

\section{Deflection angle in the strong deflection limit}
\label{secIII}
\begin{figure}[htp]
\begin{center}
    \includegraphics[width=16cm]{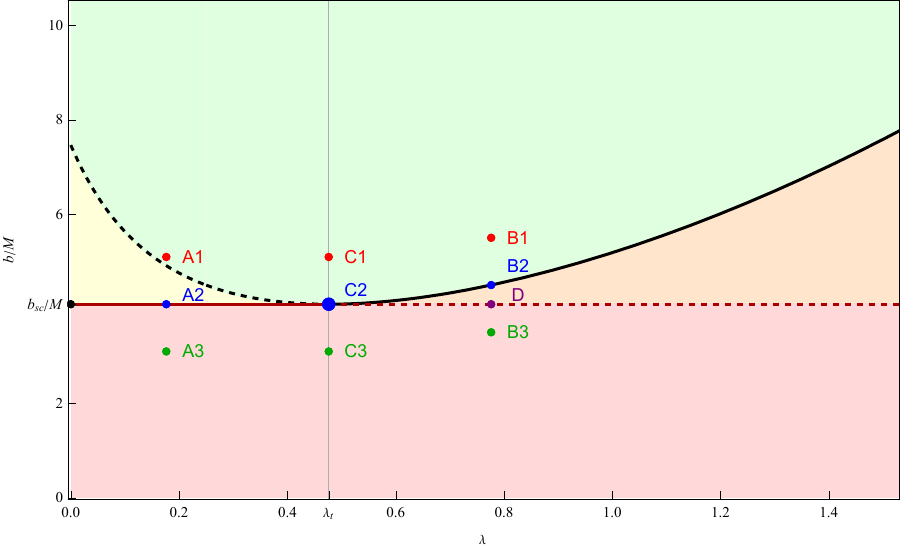}
    \caption{
    Illustration of diagram of parameter space $(\lambda,b)$, where the relevant roots of our study are the three outermost roots $r_3$, $r_4$ and $r_{\rm th}$ for $a/M=0.5$ and $s=+1$.
    The light green, orange, yellow, and red regions correspond, respectively, to the cases $r_4 > r_{\rm th} >r_3$, $r_{\rm th} > r_4 >r_3$, $r_4 > r_3 > r_{\rm th}$, and $r_3=r_4^*$ and $r_{\rm th}$.
    The red solid (dashed) line is for the double root of $r_3=r_4>r_{\rm th}$ ($r_3=r_4<r_{\rm th}$) with $\lambda < \lambda_t$ and $b= b_{sc}$ ($\lambda > \lambda_t$ and $b= b_{sc}$).
    The black solid (dashed) line is for the double root of $r_4=r_{\rm th}>r_3$ ($r_4>r_3=r_{\rm th}$) with $\lambda>\lambda_t$ and $b=b_{\rm th}$ ($\lambda<\lambda_t$ and $b=b_{\rm th}$).
    Note that $\lambda_t$ is defined by $r_{\rm th}(\lambda_t) = r_{sc}$ in (\ref{lambda0}) for a specific wormhole spin $a$, and point C2 is for the triple root.
    The points A1, A2, and A3 are along the line of $\lambda < \lambda_t$ in Sec. \ref{double_1} with the respective roots of $r_4>r_{\rm th}>r_3$,\, $r_3=r_4>r_{\rm th}$ and $r_3=r_4^*$ and $r_{\rm th}$.
    Furthermore, the point at $\lambda=0$ corresponds to the case of the black hole in \cite{hsieh-2021A}.
    The points C1, C2, and C3 are along the line of $\lambda=\lambda_t$ in Sec. \ref{triple} with the respective roots of $r_4>r_{\rm th}>r_3$,\, $r_3=r_4=r_{\rm th}$, and $r_3=r_4^*$ and $r_{\rm th}$.
    The points B1, B2, B3, and D are along the line of $\lambda>\lambda_t$ in Sec. \ref{double_2} and \ref{throat_outside} with the respective roots of $r_4>r_{\rm th}>r_3$,\, $r_4=r_{\rm th}>r_3$,\, $r_3=r_4^*$ and $r_{\rm th}$, and $r_{\rm th}>r_4=r_3$.
    We mainly consider the strong field limit (SDL).
    Nevertheless, the parameters in light green and yellow regions can lead to the light deflection in the weak field limit (WFL) in Sec. \ref{secIV}.}
    \label{parameter_space}
\end{center}
\end{figure}

In general, the impact parameter of the incident light ray $b(r_0)$ can be arbitrary.
As $r_0$ gets closer to a double or a triple root of the radial potential $\tilde R(r)$, defined as the critical distance with the corresponding impact parameter $b_{c}$, the deflection angle increases and even diverges.
In the case of black holes, it is known that the impact parameter is restricted to $b>b_c$ determined by the double root of two outermost equal roots $r_3=r_4$ \cite{chandrasekhar-1998, hsiao-2020}.
Otherwise, when $b<b_c$, the light rays will be pulled into black holes.
In wormholes with the throat, apart from the double root $r_3=r_4 \equiv r_{sc}$, given by $\tilde{R}=0$, $\tilde R'=0$ due to $R(r_{sc})=0$, $R'(r_{sc})=0$ in (\ref{R_tilde}), there might exist another double root due to $\hat \Delta(r_{\rm th})=0$, $R(r_{\rm th})=0$, namely $r_4=r_{\rm th}$ shown in Fig. \ref{parameter_space}.
Two double roots might merge to form a triple root $r_3=r_4=r_{\rm th} \equiv r_t$, obtained from $\tilde{R}(r_t)=0$, $\tilde R'(r_t)=0$, and $\tilde R''(r_t)=0$ due to $\hat \Delta(r_t)=0$, $R(r_t)=0$, and $R'(r_t)=0$, say, at $\lambda_t$ and $b (r_t)=b_t$ in (\ref{lambda0}) and (\ref{br0_k}) also shown in Fig. \ref{parameter_space}.
%

\begin{figure}[htp]
\begin{center}
    \includegraphics[width=16cm]{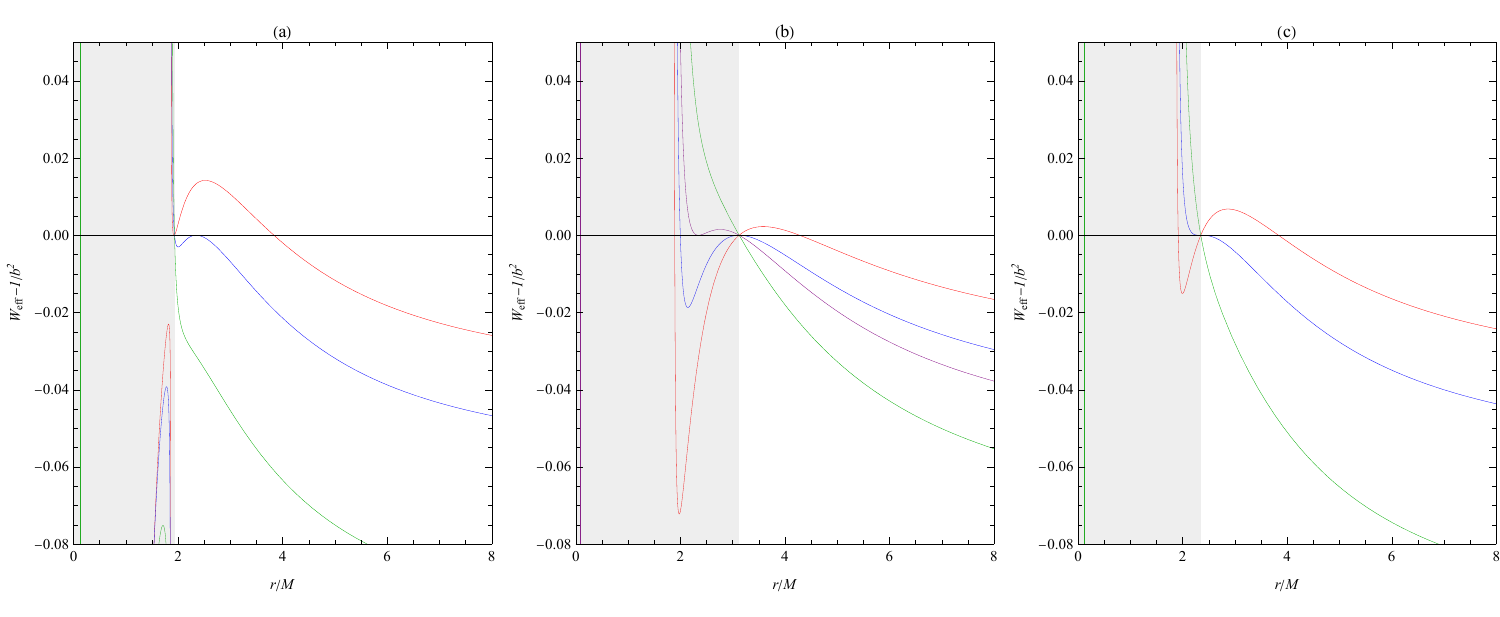}
    \caption{
    Dimensionless effective potential $W_\text{eff}(r)-1/b^2$ as a function of the radial distance $r$ with the parameters $b$ and $\lambda$ in Fig. \ref{parameter_space}.
    (a): the parameters at points A1, A2, and A3;
    (b): the parameters at points B1, B2, B3, and D;
    (c): the parameters at points C1, C2, and C3.
    The shaded area is inside the throat for reference.}
    \label{effW}
\end{center}
\end{figure}

\begin{figure}[htp]
\begin{center}
    \includegraphics[width=16cm]{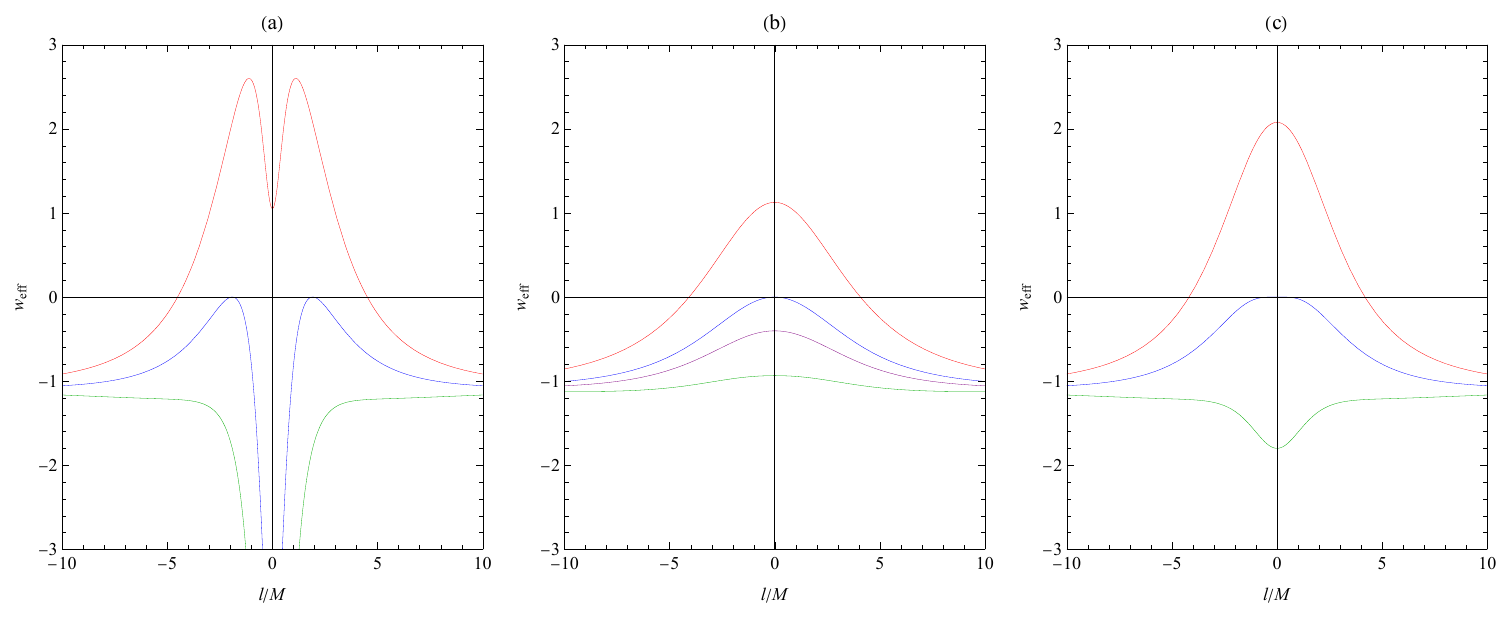}
    \caption{
    Dimensionless effective potential $w_\text{eff}(l)$ as a function of the proper radial distance $l$ with the parameters $b$ and $\lambda$ in Fig. \ref{parameter_space}.
    (a): the parameters at points A1, A2, and A3;
    (b): the parameters at points B1, B2, B3, and D;
    (c): the parameters at points C1, C2, and C3.
    Note that the incident light ray has an effective zero energy due to (\ref{w_eff}).}
    \label{effV}
\end{center}
\end{figure}

Here we give a summary of the main results of the angle of deflection as various double and triple roots are approached by the closest distance of the incident light ray around a wormhole in the SDL with the illustrative parameters marked in Fig. \ref{parameter_space}.
We emphasize the usefulness of the obtained divergence behavior of the SDL deflection angle for the analytical determination of the angular position of the relativistic images to be directly observed.
They could also serve as a unique feature of the light deflection by the throat of wormholes.
Let us start from the double root $r_3=r_4=r_{sc}$ with the critical impact parameter $b_c(r_{sc})=b_{sc}$ given by (\ref{br0_k}).
With a value $\lambda < \lambda_t$ and for $b>b_{sc}$, say at A1 in the parameter diagram, the light rays coming from spatial infinity meet the turning point at $r_0=r_4$ and return to where they started.
For $b=b_{sc}$ at A2, the light rays can undergo the unstable circular motion at $r_0=r_4=r_3$ of the double root.
However, for $b<b_{sc}$ at A3, there is no turning point outside the throat, because $r_3$ and $r_4$ are complex-conjugate roots, where the light rays pass through the throat to another spacetime to be observed.
The trajectories of light rays above can be clearly understood by the graphs of the effective potential $W_\text{eff}(r)-1/b^2$ and $w_\text{eff}(l)$ in Figs. \ref{effW}(a) and \ref{effV}(a), respectively.
The deflection angle as $b$ approaches $b_c$ with $b_c=b_{sc}$ from A1 to A2 has an approximate form as
\begin{equation} \label{hatalpha_as}
\hat{\alpha}(b) =-\bar{a}\log{\left(\frac{b}{b_{c}}-1\right)}+\bar{b} +\mathcal{O}((b-b_c) \log(b-b_c))
\end{equation}
where two parameters $\bar a$ and $\bar b$ depend on the wormhole parameters in (\ref{bar_a_b>bsc}) and (\ref{b_D_b>bsc}).
This approximate expression shows a good agreement in the SDL with the numerical calculation of the integral in (\ref{deflection_angle}) in Fig. \ref{def_angle_A}.
In contrary, for $b<b_{sc}$, the incident light rays start from the specatime of the light source at $r\to\infty$ and reach the throat at $r_{th}$ from $l\to-\infty$ to $l=0$.
Then they pass through the throat into another spacetime of the observer at spatial infinity from $l=0$ to $l\to\infty$.
%
The deflection angle with the impact parameter $b$, as approaching its critical value, from A3 to A2 in the SDL can be written in the form as
\begin{equation} \label{hatalpha_as_below}
\hat{\alpha}(b) =-\bar{a}\log{\left(\frac{b_c}{b}-1\right)}+\bar{b} + \mathcal{O}((b_c-b) \log(b_c-b))\, ,
\end{equation}
which with $\bar a$ and $\bar b$ in (\ref{bar_a_b<bsc}) and (\ref{b_D_b<bsc}) also shows a good agreement with the numerical calculation of (\ref{deflection_angle}) in Fig. \ref{def_angle_A}.

Another double root, which is also unique for wormholes, is $r_4=r_{\rm th}$ with the critical impact parameter $ b_c=b(r_{\rm th})=b_{\rm th}$ by (\ref{br0_k}) in the case of $\lambda >\lambda_t$, seen in Fig. \ref{parameter_space}.
For $ b> b_{\rm th}$ with the parameters at B1 where the roots follow $r_4 > r_{\rm th}> r_3$, the root $r_4$ serves as a turning point of the incident light ray, as seen in the plots of the effective potentials in Figs. \ref{effW}(b) and \ref{effV}(b).
As $b$ approaches $b_{\rm th}$ from B1 toward B2, the turning point shifts to the double root of $r_4=r_{\rm th}$, leading to the deflection angle of the divergent form as in (\ref{hatalpha_as}) with the critical impact parameter $b_c=b_{\rm th}$.
The coefficients $\bar a$ and $\bar b$ are obtained in (\ref{bar_a>th}) and (\ref{bar_b>th}).
For $b < b_{\rm th}$ with a value of $\lambda >\lambda_t$ where the roots are $r_{\rm th}> r_4>r_3$, there is no turning point outside the throat.
Therefore, the light rays reach the throat $r_{\rm th}$, the closest distance of the orbit, and pass through it to another spacetime to be observed.
In this case, the SDL deflection angle as $b \to b_c =b_{\rm th}$ from the below of $b_{\rm th}$ is given in the same form (\ref{hatalpha_as_below}) with $\bar a$ and $\bar b$ in (\ref{bar_a<th}) and (\ref{bar_b<th}).
The analytically approximate solution in this SDL agrees well with the numerical calculation of (\ref{deflection_angle}) in Fig. \ref{def_angle_B}.
%

Along the line of the fixed $\lambda=\lambda_t$ from C1 with $b >b_t$ to C2 with $b=b_t$, a triple root, $r_3 =r_4 =r_{\rm th} =r_t$, is met with the power-law divergence when $b \to b_c=b_t$ in the SDL obtained as \cite{tsukamoto-2020}
\be\label{hatalpha_triple}
\hat{\alpha}(b) = \bar{a} \left(\frac{b}{{b}_{c}}-1\right)^{-1/4} + \bar{b} + \mathcal{O}(b-b_c)^{3/4} \, ,\\
\ee
where $\bar a$ and $\bar b$ are obtained in (\ref{bar_a_triple_b>bc}) and (\ref{bar_b_triple_b>bc}).
For $b<b_t$ at C3 to $b=b_t$ at C2, the light rays will pass through the throat to another spacetime due to the absence of the turning point outside the throat, also seen from the graphs of the effective potentials in Figs. \ref{effW}(c) and \ref{effV}(c).
The deflection angle in the SDL can be written, when the critical impact parameter $b_t$ given by the triple root $r_t$ is approached by the impact parameter $b$, as
\be \label{hatalpha_triple_below}
\hat{\alpha}(b) = \bar{a} \left(\frac{b_{c}}{b}-1\right)^{-1/4} + \bar{b} + \mathcal{O}(b_c-b)^{3/4} \, ,
\ee
with $\bar a$ in (\ref{bar_a_triple_b<bc}) and $\bar b$ in (\ref{bar_b_triple_b<bc}).
A good agreement in the SDL is seen as compared to the numerical result of (\ref{deflection_angle}) shown in Fig. \ref{def_angle_C}.
Notice that all of the above analytical forms of the angle of deflection in the SDL play an important role as input to the analytical determination of the angular position of the relativistic images given by (\ref{theta_sn_d}), (\ref{image_a_double}), (\ref{image_s_triple}), and (\ref{image_a_triple}) to be directly observed.
Later, we will show the detailed studies of various root cases with the help of the parameters marked in Fig. \ref{parameter_space} below. 
%

\subsection{The throat inside the light sphere: $r_{\rm th} <r_4 =r_3 \equiv r_{sc}$ with the impact parameter $b_{sc}$ at point A2}
\label{double_1}
Let us now consider the unstable double root $r_3=r_4=r_{sc}$ of the radial potential $\tilde R(r)$ with a critical impact parameter $b(r_{sc}) \equiv b_{ sc}$, forming the light sphere, where the roots follow $r_4 =r_3 >r_{\rm th}$ together with the parameter $\lambda < \lambda_t$. The value of $\lambda_t$ is determined by $r_{\rm th}(\lambda_t) = r_{sc}$ for a triple root to be discussed later and is obtained as
\be \label{lambda0}
\lambda_t =\sqrt{\frac{r_{sc} (r_{sc}-2 M) +a^2}{2M r_{sc}} }.
\ee
The radius $r_{sc}$ of the light sphere is given by \cite{chandrasekhar-1998,hsiao-2020},
\be \label{rsc_k}
r_{sc} = 2 M \left\{ 1+ \cos\bigg[\frac{2}{3} \cos^{-1} \bigg( \frac{-sa}{M} \bigg) \bigg] \right\} \\
\ee
and
\be \label{bsc_k}
b_{sc} =b(r_{sc}) = -s a+ 6 M \cos\bigg[\frac{1}{3} \cos^{-1} \bigg( \frac{-sa}{M} \bigg) \bigg] .\\
\ee
The approach of the double root will be discussed from $b>b_{sc}$ (from A1 to A2) and from $b<b_{sc}$ (from A3 to A2).

\subsubsection{The case of $b>b_{sc}$ in the SDL and the approach from A1 to A2}
We now consider the impact parameter $b>b_{sc}$ in the region of $r_4 > r_3 >r_{\rm th}$ with a turning point $r_0=r_4$.
Therefore, the light rays from spatial infinity will scatter off at $r_4$ and do not reach the throat.
The deflection of the light rays is similar to that by a Kerr black hole but with modification of the coefficients $\bar a$ and $\bar b$ in the SDL due to the throat when the impact parameter approaches $b=b_{sc}$ from A1 to A2.
Let us introduce the deflection angle
\be \label{alpha}
\hat{\alpha}+\pi = 2 \int_{r_0}^{\infty} \frac{b r(r-2M)+2Msar}{\sqrt{(r-r_+)(r-r_-)(r-r_{\rm th})(r-\tilde{r}_-) (r-r_1)(r-r_2)(r-r_3)(r-r_0)}} dr.
\ee
To do the integration, we change the variable to $z$
\be\label{z}
z \equiv 1-\frac{r_0}{r}
\ee
and the integral can be rewritten as
\be \label{alpha_r0}
\hat{\alpha} +\pi = 2r_0^2 \int_{0}^{1} \frac{[r_0+2M(z-1)]b -2Msa(z-1)}{\sqrt{ B(z,r_0)}} dz,
\ee
where the denominator can be obtained straightforwardly by the change of the variable as
\be
\begin{split} \label{B}
B(z,r_0)=& r_0 [r_0+r_+ (z-1)] [r_0+r_- (z-1)] [r_0+r_{\rm th} (z-1)] [r_0+\tilde{r}_{-} (z-1)] \\
& \times [r_0+r_1 (z-1)] [r_0+r_2 (z-1)] [r_0+r_3 (z-1)] z \\
=& \sum_{n=1}^8 c_n(r_0) z^n
\end{split}
\ee
with the coefficients $c_n$'s.
In particular, the coefficients of relevance to the SDL are $c_1 (r_0)$ and $c_2 (r_0)$ given by
\be \label{c1_A>}
\begin{split}
c_1(r_0) =& r_0 (r_0-r_+)(r_0-r_-)(r_0-r_{\rm th})(r_0-\tilde{r}_-) (r_0-r_1)(r_0-r_2)(r_0-r_3),\\
c_2(r_0) =& \frac{1}{2} \bigg\{ 2 (r_0-r_+) (r_0-r_{\rm th}) (r_0-\tilde{r}_-) \bigg[r_0 r_1 (r_0-r_2) (r_0-r_3) (r_0-r_-)\\
&+r_0 r_2 (r_0-r_1) (r_0-r_3) (r_0-r_-)+r_0 r_3 (r_0-r_1) (r_0-r_2) (r_0-r_-)\\
&+r_0 r_- (r_0-r_1) (r_0-r_2) (r_0-r_3) \bigg]\\
&+r_0 (r_0-r_1) (r_0-r_2) (r_0-r_3) (r_0-r_-) \bigg[2 (r_0-\tilde{r}_-) (r_+ (r_0-r_{\rm th})+r_{\rm th} (r_0-r_+)) \\
&+2 \tilde{r}_- (r_0-r_+) (r_0-r_{\rm th}) \bigg] \bigg\},
\end{split}
\ee
where $c_1(r_0)$ vanishes as $r_0 \to r_{sc}=r_3=r_4$ while other coefficients remain nonzero.
Now the deflection angle is a function of the closest approach distance $r_0$ given by (\ref{deflection_angle}) as
\be
\hat\alpha(r_0) = I (r_0) -\pi \, , \quad I(r_0)= \int_0^1 f(z,r_0) dz \,
\ee
with the integrand function
\be
f(z,r_0)= \frac{2r_0^2 [r_0 b +2Mb(z-1) -2Msa(z-1)]}{\sqrt{c_1 z +c_2 z^2 +c_3 z^3 +c_4 z^4 +c_5 z^5 +c_6 z^6 +c_7 z^7 +c_8 z^8}} \,.
\ee
In the limit $r_0 \to r_{sc}$ of the double root $r_3=r_4$, the coefficients
\be
\begin{split}
c_1(r_0) &\sim \mathcal{O}(r_0-r_{sc}) \to 0,
\end{split}
\ee
and {$c_n(r_{sc})$ for $n>1$ is nonzero}.
As a result, the integrand function behaves $f(z,r_0) \to 1/z$ for small $z$, leading to the divergence.
We then separate the divergence piece from the finite one, where the divergence can be extracted from the integral of the function
\be
f_D(z,r_0) \equiv \frac{2r_0^2 [r_0 b +2Mb (z-1) -2Msa(z-1)]}{\sqrt{c_1 z +c_2 z^2}} \, ,
\ee
and the integration of
\be \label{f_R}
f_R(z,r_0) =f(z,r_0) -f_D(z,r_0)
\ee
over $z$ is finite.
We proceed by considering the divergent part,
\be
\begin{split} \label{I_D_r0}
I_D(r_0)
=& \int_0^1 f_D(z,r_0) dz\\
=& -\frac{4 M r_0^2 (sa-b) \sqrt{c_1 +c_2}}{c_2} \\
&+ \frac{4r_0^2 [M sa (c_1+2 c_2)-M b c_1 +b (r_0-2 M) c_2]}{c_2^{3/2}} \log{\left( \frac{\sqrt{c_1+c_2} -\sqrt{c_2}}{\sqrt{c_1}} \right)}\, . \\
\end{split}
\ee
In the SDL, the expansions of the coefficient $c_1(r_0)$ and the impact parameter $b(r_0)$ in powers of small $r_0-r_{sc}$ read
\be \label{c1_expand}
c_1(r_0)= c_{1 sc}' (r_0-r_{sc}) +\mathcal{O}(r_0-r_{sc})^2 \,
\ee
and
\be \label{b_r_0}
b(r_0) = b_{sc} +\frac{1}{2!} b_{sc}'' (r_0-r_{sc})^2 +\mathcal{O}(r_0-r_{sc})^3 ,
\ee
where $c_{1 sc}' \equiv c_1'(r_{sc})$ and $b_{sc}'' \equiv b''(r_{sc})$, and the prime means the derivative with respect to $r_0$.
Therefore, in the limit of $r_0\to r_{sc}$, the element in the logarithm can be expanded as
\be \label{in_log}
\lim_{r_0\to r_{sc}} \frac{\sqrt{c_1(r_0)+c_2(r_0)}-\sqrt{c_2(r_0)}}{\sqrt{c_1(r_0)}}
= \frac{\sqrt{c_1'(r_{sc}) }}{2 \sqrt{c_2(r_{sc})}} \left(r_0 -r_{sc} \right)^{1/2}.
\ee
Plugging (\ref{c1_expand}), (\ref{b_r_0}), and (\ref{in_log}) into the divergent part $I_D$ gives
\be
\begin{split}
I_D(r_0) \simeq & -\frac{4 M r_{sc}^2 (sa-b_{sc}) }{\sqrt{c_{2 sc}}} \\
&- \frac{2r_{sc}^2 [2 M sa +b_{sc} (r_{sc}-2 M)]}{\sqrt{c_{2 sc}}} \log{\left[ \frac{c_{1 sc}' }{4 c_{2 sc}} (r_0-r_{sc}) \right]}\, .
\end{split}
\ee
Or replacing $r_0$ by $b$ through (\ref{b_r_0}), the formula becomes
\be
\begin{split}
I_D(b) \simeq
& -\frac{r_{sc}^2 [2 M sa +b_{sc} (r_{sc}-2 M)]}{\sqrt{c_{2 sc}}} \log{\left( \frac{b}{b_{sc}}-1 \right)} \\
&- \frac{2r_{sc}^2 [2 M sa +b_{sc} (r_{sc}-2 M)]}{\sqrt{c_{2 sc}}} \log{\left( \frac{c_{1 sc}' }{4 c_{2 sc}} \sqrt{\frac{2 b_{sc}}{b_{sc}''}} \right)}
-\frac{4 M r_{sc}^2 (sa-b_{sc}) }{\sqrt{c_{2 sc}}} \\
\end{split}
\ee
through the limit of
\be \label{r_0_b_s}
\lim_{r_0\to r_{sc}} (r_0 -r_{sc}) = \lim_{b \to b_{sc}} \sqrt{\frac{2}{b_{sc}''} (b -b_{sc})}\, .
\ee
Finally, the coefficients $\bar{a}$ and $b_D$ for the divergent part are
\be
\bar{a} = \frac{r_{sc}^2 [2 M sa +b_{sc} (r_{sc}-2 M)]}{\sqrt{c_{2 sc}}}\label{bar_a_b>bsc}
\ee
and
\be
b_D = -2\bar{a} \log{\left( \frac{c_{1 sc}' }{4 c_{2 sc}} \sqrt{\frac{2 b_{sc}}{b_{sc}''}} \right)}
-\frac{4 M r_{sc}^2 (sa-b_{sc}) }{\sqrt{c_{2 sc}}}\, . \label{b_D_b>bsc}
\ee

Here, we provide another way to verify the coefficients of the divergent part.
We first take the strong deflection limit such that $c_1(r_{sc})=0$.
Then, we can expect the main contribution of the integral, which comes from the $z \to 0$ region.
The natural lower limit can be set at $z \to {c_1}/{c_2} ={2}(r_0-r_{sc})/{r_{sc}}$ when $c_1 z \simeq c_2 z^2$ for $r_0 \to r_{sc}$.
So, the divergent part can be found to be
\be
\begin{split}
I_D(r_0) \simeq & \int_0^1 \lim_{z\to 0} f_D(z,r_{sc}) dz\\
       = & \frac{2r_{sc}^2 [r_{sc} b_{sc} -2Mb_{sc} +2Msa] }{\sqrt{c_{2 sc}}} \int_{{z \rightarrow \frac{2}{r_{sc}} (r_0-r_{sc})}}\frac{1}{z} dz + {\rm finite \, part} \\
       \sim & - \frac{2r_{sc}^2 [r_{sc} b_{sc} -2Mb_{sc} +2Msa] }{\sqrt{c_{2 sc}}} \log{( r_0-r_{sc})} + {\rm finite \, part} \, . \\
\end{split}
\ee
Or we can also rewrite $I_D$ as a function of the impact parameter through (\ref{r_0_b_s}) as
\be
I_D(b) \sim - \frac{r_{sc}^2 [2Msa +b_{sc} (r_{sc}-2M)]}{\sqrt{c_{2 sc}}} \log{\left( \frac{b}{b_{sc}}-1 \right)} \, ,
\ee
given the same $\bar{a}$ in (\ref{bar_a_b>bsc}).

Moreover, we would like to ask whether the coefficient $\bar{a}$ due to a Kerr-like wormhole can be reduced to that due to a Kerr black hole in \cite{hsieh-2021A, hsieh-2021B}.
By turning off the parameter $\lambda$, the throat $r_{\rm th}$ reduces to $r_+$ and $\tilde{r}_-=r_-$.
{The coefficient $c_{2 sc}$ in (\ref{c1_A>})} becomes
\be
c_{2 sc} = r_{sc}^2 (r_{sc}-r_+)^2(r_{sc}-r_-)^2 (r_{sc}-r_1)(r_{sc}-r_2)
\ee
with the root $r_1=-2 r_{sc}$ resulting from $r_2=0$ and the root property $r_1 +r_2 +r_3 +r_4 =0$ in \cite{wang-2022}.
Then, together with the condition of the light sphere \cite{chandrasekhar-1998}
\begin{equation}
r_{sc}^2 -3M r_{sc} +2sa \sqrt{M r_{sc}}=0 ,
\end{equation}
{and $b_{sc}=b(r_{sc})$ in (\ref{br0_k})},
the coefficient in (\ref{bar_a_b>bsc}) becomes
\be
\left. \bar{a} \right|_{\lambda =0} =\frac{2M r_{sc}}{(r_{sc}-M) \sqrt{3M r_{sc}}} ,
\ee
which is $\bar a$ due to a Kerr black hole \cite{hsieh-2021A}.

The subleading term is obtained from the integration of $f_R(z,r_{sc})$ in (\ref{f_R}) denoted by $b_R$,
\be \label{b_R}
b_R = I_R(r_{sc}) = \int_0^1 f_R(z,r_{sc}) dz ,\\
\ee
which is a constant to be computed numerically.
Thus, the coefficient $\bar{b}$ can be computed from the sum of $b_D$ and $b_R$
\be\label{bar_b}
\bar{b}=-\pi+b_D+b_R.
\ee

\begin{figure}[htp]
\begin{center}
    \includegraphics[width=13cm]{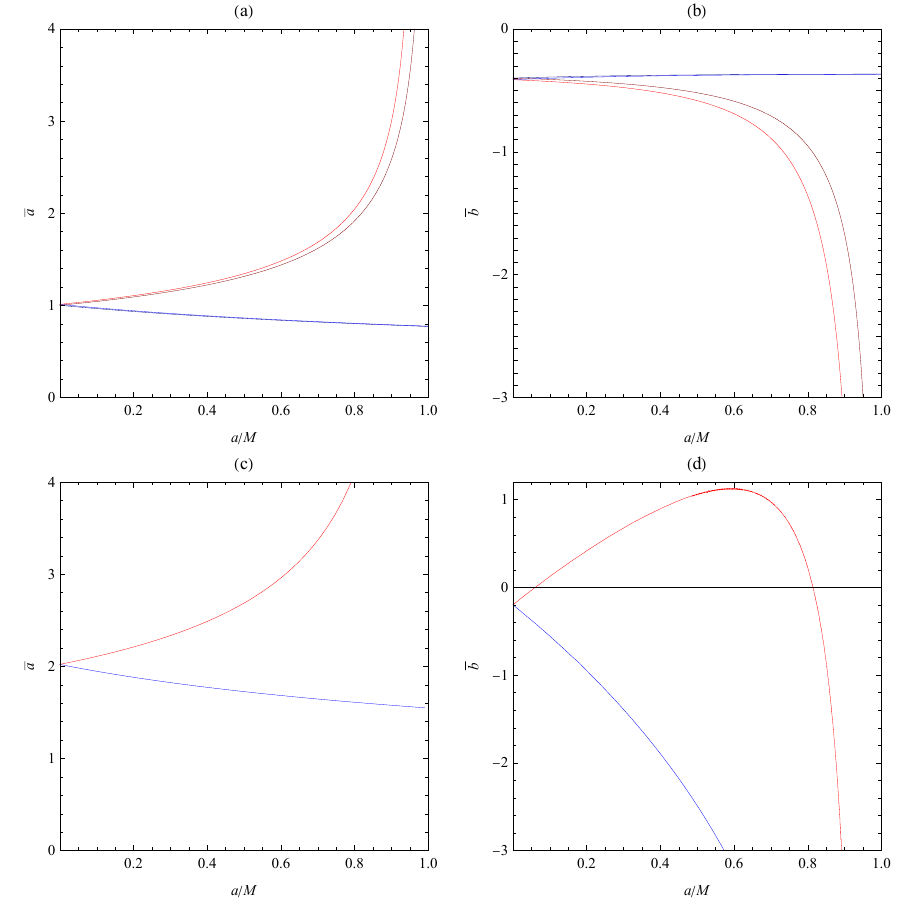}
    \caption{
    (a): the coefficients $\bar{a}$ as the functions of the spin $a$ with (\ref{bar_a_b>bsc}) for $ b> b_{sc}$;
    (b): the coefficients $\bar{b}$ as the functions of the spin $a$ mainly with (\ref{b_D_b>bsc}) from $ b> b_{sc}$, for a direct orbit in the red (deep red) line and a retrograde orbit in the blue (deep blue) line for $\lambda=0$ ($\lambda=0.1 $);
    (c): the coefficients $\bar{a}$ as the functions of the spin $a$ with (\ref{bar_a_b<bsc}) for $ b<b_{sc}$;
    (d): the coefficients $\bar{b}$ as the functions of the spin $a$ mainly with (\ref{b_D_b<bsc}) for $b<b_{sc}$ for a direct orbit in the red line and a retrograde orbit in the blue line for $\lambda=0.1 $.
    Note that, the SDL deflection angle with $b<b_{sc}$ cannot be reduced to the Kerr black hole case since the light ray cannot travel through a black hole.}
    \label{barab_A}
\end{center}
\end{figure}

\subsubsection{The case of $b<b_{sc}$ in the SDL and the approach from A3 to A2}
We turn to an alternative approach from A3 to A2 and start considering $b<b_{sc}$ in the region of the roots $r_4=r_3^*$ of a pair of complex conjugates in Fig. \ref{parameter_space}.
Because there is no turning point outside the throat, light rays will pass through the throat.
In this case, the deflection angle (\ref{alpha}) in terms of $z$ defined as
\begin{equation}
z=1-\frac{r_{sc}}{r}
\end{equation}
with $r_3 =r_4=r_{sc}$ becomes
\be
\hat{\alpha}+\pi = 2r_{sc}^2 \int_{1-\frac{r_{sc}}{r_{\rm th}}}^{1} \frac{[r_{sc}+2M(z-1)]b -2Msa(z-1)}{\sqrt{ \mathcal{B}(z,b)}} dz ,
\ee
where the denominator is
\be
\begin{split}
\mathcal{B}(z,b) =& [r_{sc}+r_+ (z-1)] [r_{sc}+r_- (z-1)] [r_{sc}+r_{\rm th} (z-1)] [r_{sc}+\tilde{r}_{-} (z-1)] \\
&\times [r_{sc}+r_1 (z-1)] [r_{sc}+r_2 (z-1)] [r_{sc}+r_3 (z-1)] [r_{sc}+r_4 (z-1)] \\
=& \sum_{n=0}^8 c_n(b) z^n \, .
\end{split}
\ee
{The coefficients $c_0$ and $c_1$ involved in the SDL are
\be
\begin{split}
c_0(b) =& (r_{sc}-r_+)(r_{sc}-r_-)(r_{sc}-r_{\rm th})(r_{sc}-\tilde{r}_-) (r_{sc}-r_1)(r_{sc}-r_2)(r_{sc}-r_3) (r_{sc}-r_4) ,\\
c_1(b) =& r_+ (r_{sc}-r_1) (r_{sc}-r_2) (r_{sc}-r_3) (r_{sc}-r_4) (r_{sc}-r_-) (r_{sc}-r_{\rm th}) (r_{sc}-\tilde{r}_-)\\
&+r_- (r_{sc}-r_1) (r_{sc}-r_2) (r_{sc}-r_3) (r_{sc}-r_4) (r_{sc}-r_+) (r_{sc}-r_{\rm th}) (r_{sc}-\tilde{r}_-)\\
&+r_4 (r_{sc}-r_1) (r_{sc}-r_2) (r_{sc}-r_3) (r_{sc}-r_-) (r_{sc}-r_+) (r_{sc}-r_{\rm th}) (r_{sc}-\tilde{r}_-)\\
&+r_3 (r_{sc}-r_1) (r_{sc}-r_2) (r_{sc}-r_4) (r_{sc}-r_-) (r_{sc}-r_+) (r_{sc}-r_{\rm th}) (r_{sc}-\tilde{r}_-)\\
&+r_2 (r_{sc}-r_1) (r_{sc}-r_3) (r_{sc}-r_4) (r_{sc}-r_-) (r_{sc}-r_+) (r_{sc}-r_{\rm th}) (r_{sc}-\tilde{r}_-)\\
&+r_1 (r_{sc}-r_2) (r_{sc}-r_3) (r_{sc}-r_4) (r_{sc}-r_-) (r_{sc}-r_+) (r_{sc}-r_{\rm th}) (r_{sc}-\tilde{r}_-)\\
&+r_{\rm th} (r_{sc}-r_1) (r_{sc}-r_2) (r_{sc}-r_3) (r_{sc}-r_4) (r_{sc}-r_-) (r_{sc}-r_+) (r_{sc}-\tilde{r}_-)\\
&+\tilde{r}_- (r_{sc}-r_1) (r_{sc}-r_2) (r_{sc}-r_3) (r_{sc}-r_4) (r_{sc}-r_-) (r_{sc}-r_+) (r_{sc}-r_{\rm th}) .
%
\end{split}
\ee
As $b \to b_{sc}$, $r_3, r_4 \to r_{sc}$ in (\ref{r_3}) and (\ref{r_4}).
Then $c_0(b_{sc})$ and $c_1(b_{sc})$ vanish whereas $c_{n} (b_{sc})$ for $n>1$ remains nonzero.}
The integrand function, giving the divergence of the deflection angle in the case of the double root, is then given by
\be
f_D(z,b) \equiv \frac{ 2r_{sc}^2 [r_{sc}b+2Mb(z-1) -2Msa(z-1)]}{\sqrt{ c_0 +c_1 z +c_2 z^2}}.
\ee
After the integration over $z$, $I_D(b)$ is obtained as
\begin{eqnarray}
I_D(b) =&&
-\frac{4 r_{sc}^2 M (sa-b) (1-\frac{r_{sc}-r_{\rm th}}{r_{\rm th}})}{\sqrt{c_2(b)}}
+\frac{2 r_{sc}^2 [2M sa+b (r_{sc}-2 M)] }{\sqrt{c_2(b)}} \log{\left( \frac{\sqrt{c_0+c_1+c_2}-\sqrt{c_0}+\sqrt{c_2}}{\sqrt{c_0+c_1+c_2}-\sqrt{c_0}-\sqrt{c_2}} \right)}\nonumber\\
&& \scalebox{1.2}{$ -\frac{2 r_{sc}^2 [2M sa +b (r_{sc}-2 M)] }{\sqrt{c_2(b)}}
\log{\left[ \frac{\sqrt{r_{\rm th} (c_0 r_{\rm th}-c_1 r_{sc}+c_1 r_{\rm th})+c_2 (r_{sc}-r_{\rm th})^2} -\sqrt{c_0} r_{\rm th}-\sqrt{c_2} (r_{sc}-r_{\rm th})}{\sqrt{r_{\rm th} (c_0 r_{\rm th}-c_1 r_{sc}+c_1 r_{\rm th})+c_2 (r_{sc}-r_{\rm th})^2} -\sqrt{c_0} r_{\rm th}+\sqrt{c_2} (r_{sc}-r_{\rm th})} \right]} $}\, .\nonumber\\
\end{eqnarray}
Hence, we expand the coefficients $c_0(b)$ and $c_1(b)$ around $b=b_{sc}$ as
\be
\begin{split}
c_0(b) &= c_{0}'(b_{sc}) (b-b_{sc}) +\mathcal{O}(b-b_{sc})^2\, , \\
c_1(b) &= c_{1}'(b_{sc}) (b-b_{sc}) +\mathcal{O}(b-b_{sc})^2\, , \\
\end{split}
\ee
where the prime means the derivative with respect to $b$.
Note that $c_0'(b_{sc})$, $c_1'(b_{sc})<0$, while other $c_n(b_{sc})$ for $n >1$ is nonzero.
So, the integral $I_D(b)$ becomes
\be
\begin{split}
I_D(b) \simeq &
-\frac{2 r_{sc}^2 [2M sa +b_{sc} (r_{sc}-2 M)]}{\sqrt{c_2(b_{sc})}} \log{\left(\frac{b_{sc}}{b} -1\right)} \\
& -\frac{2 r_{sc}^2 [2M sa +b_{sc} (r_{sc}-2 M)]}{\sqrt{c_2(b_{sc})}} \log{\left( \frac{-b_{sc} c_0'(b_{sc}) }{4 c_{2}(b_{sc}) } \frac{r_{\rm th}}{r_{sc}-r_{\rm th}} \right)}
-\frac{4M r_{sc}^2 (sa-b_{sc}) (2 r_{\rm th} -r_{sc})}{\sqrt{c_{2}(b_{sc})} r_{\rm th}} ,\\
\end{split}
\ee
and the coefficients $\bar{a}$ and $b_D$ are obtained as
\be
\bar{a} =\frac{2 r_{sc}^2 [2M sa+b_{sc} (r_{sc}-2 M)]}{\sqrt{c_{2}(b_{sc})} }
\label{bar_a_b<bsc}
\ee
and
\be
b_D = -\bar{a} \log{\left( \frac{-b_{sc} c_{0}'(b_{sc}) }{4 c_{2}(b_{sc}) } \frac{r_{\rm th}}{r_{sc}-r_{\rm th}} \right)}
-\frac{4M r_{sc}^2 (sa-b_{sc}) (2 r_{\rm th} -r_{sc})}{\sqrt{c_{2}(b_{sc})} r_{\rm th}} \, .\label{b_D_b<bsc}
\ee
The value of $b_R$ can be computed from the integral of $f_R(z,b_{sc})$ to be done numerically.
Thus, the coefficient $\bar{b}$ can be calculated from the sum of $b_D$ and $b_R$ in (\ref{bar_b}).

\begin{figure}[htp]
\begin{center}
    \includegraphics[width=15cm]{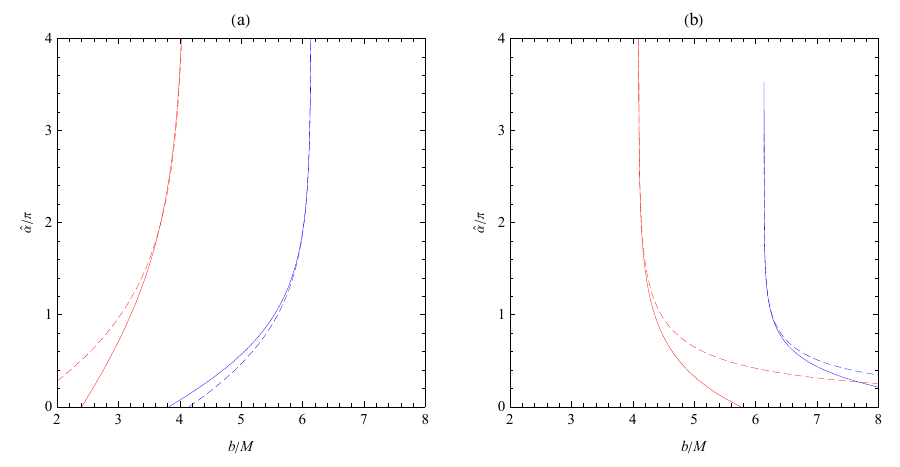}
    \caption{
    The SDL deflection angle $\hat{\alpha}(b)$ for $b<b_{sc}$ in (a) with (\ref{hatalpha_as_below}), (\ref{bar_a_b<bsc}), and (\ref{bar_b}); for $b>b_{sc}$ in (b) with (\ref{hatalpha_as}), (\ref{bar_a_b>bsc}), and (\ref{bar_b}), for a direct orbit (red line) and a retrograde orbit (blue line) with $a/M=0.5$ and $\lambda=0.3$.
    The numerical results obtained from calculating the integral of (\ref{deflection_angle}) are displayed in dashed lines for comparison.}
    \label{def_angle_A}
\end{center}
\end{figure}

Notice that in general $\bar a$ and $\bar b$ in (\ref{bar_a_b<bsc}) and (\ref{b_D_b<bsc}) from the approach of $b< b_{sc}$ are different from those in (\ref{bar_a_b>bsc}) and (\ref{b_D_b>bsc}) from $b >b_{sc}$.
A factor 2 difference in $\bar a$ between (\ref{bar_a_b<bsc}) and (\ref{bar_a_b>bsc}) can be attributed to the fact that from $b > b_{sc}$ toward $b \to b_{sc}$, the light rays will meet the turning point $r_0=r_4$ and deflect to another direction as seen in the graphs in Fig. \ref{effV}(a).
However, from $b < b_{sc}$, the light rays pass through the throat to another spacetime also seen in Fig. \ref{effV}(a).
When $b \to b_{sc}$ with two developed unstable maxima in each spacetimes, the light ray may travel between them, giving the extra factor 2 in $\bar a$ compared to the case of $b > b_{sc}$.
The coefficient $\bar b$ is different between both cases.
The coefficients $\bar a$ and $\bar b$ for a direct orbit and a retrograde orbit from both $b > b_{sc}$ and $b< b_{sc}$ toward the double root are plotted with a finite value $\lambda$ as compared with the Kerr case with $\lambda=0$ in Fig. \ref{barab_A}, where $\bar a$ and $\vert \bar b \vert$ increase in $\lambda$.
The corresponding deflection angle in the SDL is shown in Fig. \ref{def_angle_A}.
We also numerically calculate the integral of (\ref{deflection_angle}) and show a good agreement with the analytical approximate solution in the SDL.

\subsection{The throat on a light sphere: $r_{\rm th} =r_4 $ with the impact parameter $b_{\rm th}$ at point B2}
\label{double_2}
Another double root is that the throat at $r_{\rm th}$ together with $r_4$ becomes an unstable double root $r_{\rm th}=r_4$ with the parameter $\lambda > \lambda_t$, say at point B2 in Fig. \ref{parameter_space}.
The light rays are allowed to have rather unique unstable circular orbits with the radius at the throat $r_{\rm th}$, forming a light sphere \cite{kasuya-2021}.
The roots follow $r_{\rm th} =r_4 >r_3 $.
The limit of $r_{\rm th}=r_4$ can be approached either from $b>b_{\rm th} $ or from $b<b_{\rm th}$.

\subsubsection{The case of $ b>b_{\rm th}$ in the SDL and the approach from B1 to B2}
In the former case, we start from the parameters in the region of $r_4> r_{\rm th}$ and $\lambda > \lambda_t$ at point B1 with the impact parameter $b$, and then shift the parameters toward the double root at B2 with $b=b_{\rm th}$.
So, the turning point is set at $r_0=r_4$ with the impact parameter $b(r_0)$ in (\ref{br0_k}).
We introduce the variable $z$ as usual and rewrite the integration in (\ref{alpha}) as in (\ref{alpha_r0}) in the same form of $B(z,r_0)$ in (\ref{B}).
Then, the integral of the divergent part has the same result of (\ref{I_D_r0}).
In the limit $r_0 \to r_{\rm th}$, the coefficient $c_1$ in (\ref{c1_A>}) follows
\be
\begin{split}
c_1(r_0) &\sim \mathcal{O}(r_0-r_{\rm th}) \to 0,
\end{split}
\ee
{whereas the other $c_n(r_{\rm th})$ for $n>1$ is nonzero.}
We expand the coefficient $c_1(r_0)$ and the impact parameter $b(r_0)$ in terms of $r_0-r_{\rm th}$ as
\be
c_1(r_0) = c_{1 \rm th}' (r_0-r_{\rm th}) +\mathcal{O}(r_0-r_{\rm th})^2 \, ,
\ee
and
\be \label{b_r0_r0}
b(r_0) = b_{\rm th} +b_{\rm th}' (r_0-r_{\rm th}) +\mathcal{O}(r_0-r_{\rm th})^2,
\ee
where $c_{1 \rm th}' \equiv c_1'(r_{\rm th})$ and $b_{\rm th}' \equiv b'(r_{\rm th})$ are nonzero {with the different values from Sec. \ref{double_1}}.
So, when we replace $r_0$ with $b(r_0)$, the divergent part becomes
\be
\begin{split}
I_D(b) \simeq & - \frac{2r_{\rm th}^2 [2 M sa +b_{\rm th} (r_{\rm th}-2 M)]}{\sqrt{c_{2 \rm th}}} \log{\left( \frac{b}{b_{\rm th}}-1 \right)} \\
& - \frac{2r_{\rm th}^2 [2 M sa +b_{\rm th} (r_{\rm th}-2 M)]}{\sqrt{c_{2 \rm th}}} \log{\left( \frac{c_{1 \rm th}'}{4 c_{2 \rm th}} \frac{b_{\rm th}}{b_{\rm th}'} \right)}
-\frac{4 M r_{\rm th}^2 (sa-b_{\rm th}) }{\sqrt{c_{2 \rm th}}} \\
\end{split}
\ee
given by the limit of (\ref{b_r0_r0})
\be
\lim_{r_0\to r_{\rm th}} (r_0 -r_{\rm th}) = \lim_{b\to b_{\rm th}} \frac{1}{b_{\rm th}'} (b -b_{\rm th}).
\ee
Finally, the coefficients $\bar{a}$ and $b_D$ in the divergent part are
\be
\bar{a} = \frac{2r_{\rm th}^2 [2 M sa +b_{\rm th} (r_{\rm th}-2 M)]}{\sqrt{c_{2 \rm th}}} \label{bar_a>th}
\ee
and
\be
b_D = -\bar{a} \log{\left( \frac{c_{1 \rm th}' }{4 c_{2 \rm th}} \frac{b_{\rm th}}{b_{\rm th}'} \right)}
-\frac{4 M r_{\rm th}^2 (sa-b_{\rm th}) }{\sqrt{c_{2 \rm th}}},
\label{bar_b>th}
\ee
{with $c_{2 \rm th} \equiv c_2(r_{\rm th})$.}
The value of $b_R$ given by the integral of the function $f_R(z,r_{\rm th})$ can be calculated through (\ref{b_R}) by replacing $r_{sc}$ with $r_{\rm th}$.
Thus, the coefficient $\bar{b}$ can be computed from the sum of $b_D$ and $b_R$ in (\ref{bar_b}).

\begin{figure}[htp]
\begin{center}
    \includegraphics[width=15cm]{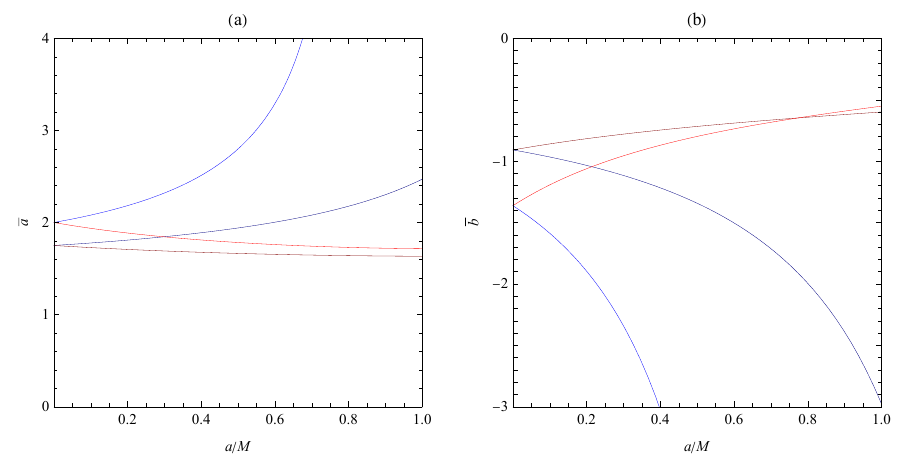}
    \caption{
    (a): the coefficients $\bar{a}$ as the functions of the spin $a$ with (\ref{bar_a>th}) for both $ b> b_{\rm th}$ and $ b< b_{\rm th}$, which has the same value;
    (b): the coefficients $\bar{b}$ as the functions of the spin $a$ mainly with (\ref{bar_b>th}) for both $ b> b_{\rm th}$ and $ b< b_{\rm th}$, which has the same value, for a direct orbit in the red (deep red) line and a retrograde orbit in the blue (deep blue) line with $\lambda=1.0$ ($\lambda=1.2 $).}
     \label{barab_B}
\end{center}
\end{figure}

\subsubsection{The case of $ b<b_{\rm th}$ in the SDL and the approach from the below of $b_{\rm th}$ to B2 }
For the unstable double root $r_{\rm th} =r_4$ approached from $b< b_{\rm th}$, we consider the impact parameter $b$ in the region of $r_{\rm th} > r_4>r_3$ and $\lambda > \lambda_t$ with the parameters in the light orange region in Fig. \ref{parameter_space}.
Then, the deflection angle (\ref{alpha}) in terms of $z$ defined to be
\begin{equation}\label{z_th}
z=1-\frac{r_{\rm th}}{r}
\end{equation}
becomes
\be \label{alpha_th}
\hat{\alpha}+\pi = 2r_{\rm th}^2 \int_{0}^{1} \frac{[r_{\rm th}+2M(z-1)]b -2Msa(z-1)}{\sqrt{ \mathcal{B}(z,b)}} dz,
\ee
where the denominator becomes
\be
\begin{split}\label{B_th}
\mathcal{B}(z,b) =& r_{\rm th} [r_{\rm th}+r_+ (z-1)] [r_{\rm th}+r_- (z-1)] [r_{\rm th}+\tilde{r}_{-} (z-1)] \\
&\times [r_{\rm th}+r_1 (z-1)] [r_{\rm th}+r_2 (z-1)] [r_{\rm th}+r_3 (z-1)] [r_{\rm th}+r_4 (z-1)] z \\
=& \sum_{n=1}^8 c_n(b) z^n \,.
\end{split}
\ee
{In particular, the coefficients $c_1$ and $c_2$ are expressed as
\be
\begin{split}\label{c_B<}
c_1(b) =& r_{\rm th} (r_{\rm th}-r_+)(r_{\rm th}-r_-)(r_{\rm th}-\tilde{r}_-) (r_{\rm th}-r_1)(r_{\rm th}-r_2)(r_{\rm th}-r_3) (r_{\rm th}-r_4), \\
c_2(b) =& \frac{1}{2} \bigg\{ 2 (r_{\rm th}-r_-) (r_{\rm th}-r_+) (r_{\rm th}-\tilde{r}_-) \bigg[r_4 r_{\rm th} (r_{\rm th}-r_1) (r_{\rm th}-r_2) (r_{\rm th}-r_3)\\
&+r_2 r_{\rm th} (r_{\rm th}-r_1) (r_{\rm th}-r_3) (r_{\rm th}-r_4)+r_1 r_{\rm th} (r_{\rm th}-r_2) (r_{\rm th}-r_3) (r_{\rm th}-r_4)\\
&+r_3 r_{\rm th} (r_{\rm th}-r_1) (r_{\rm th}-r_2) (r_{\rm th}-r_4) \bigg]\\
&+r_{\rm th} (r_{\rm th}-r_1) (r_{\rm th}-r_2) (r_{\rm th}-r_3) (r_{\rm th}-r_4) \bigg[ 2 (r_{\rm th}-\tilde{r}_-) (r_+ (r_{\rm th}-r_-)+r_- (r_{\rm th}-r_+))\\
&+2 \tilde{r}_- (r_{\rm th}-r_-) (r_{\rm th}-r_+) \bigg] \bigg\}\, ,
\end{split}
\ee
where $c_1(b)$ vanishes as $b \to b_{\rm th}$ due to $r_4 \to r_{\rm th}$ in (\ref{r_4}), while $c_n (b_{\rm th})$ for $ n>1$ is nonzero.}

Then, the integrand function that potentially leads to the divergence of the deflection angle as approaching the double root is found to be
\be
f_D(z,b) \equiv \frac{ 2r_{\rm th}^2 [r_{\rm th}b+2Mb(z-1) -2Msa(z-1)]}{\sqrt{ c_1 z +c_2 z^2}},
\ee
and after the integration over $z$, the divergent part $I_D(b)$ is obtained as
\be
I_D(b) =
-\frac{4 r_{\rm th}^2 M (sa-b)}{\sqrt{c_2(b)}}
-\frac{4 r_{\rm th}^2 [2M sa +b (r_{\rm th}-2 M)] }{\sqrt{c_2(b)}} \log{\left(\frac{\sqrt{c_1(b)+c_2(b)}-\sqrt{c_2(b)}}{\sqrt{c_1(b)}}\right)}.
\ee
We expand the coefficient $c_1(b)$ around $b=b_{\rm th}$ as
\begin{equation}
c_1(b) = c_{1}'(b_{\rm th}) (b-b_{\rm th}) +\mathcal{O}(b-b_{\rm th})^2,
\end{equation}
with $c_1'(b_{\rm th})<0$ where the element of the logarithmic function in this limit becomes
\be
\lim_{b\to b_{\rm th}} \frac{\sqrt{c_1(b)+c_2(b)}-\sqrt{c_2(b)}}{\sqrt{c_1(b)}}
= \frac{\sqrt{-c_1'(b_{\rm th}) b}}{2 \sqrt{c_2(b_{\rm th})}} \left(\frac{b_{\rm th}}{b} -1\right)^{1/2}
\ee
for $b \lesssim b_{\rm th}$.
So, $I_D(b)$ becomes
\be
\begin{split}
I_D(b) \simeq &
-\frac{2 r_{\rm th}^2 [2M sa +b_{\rm th} (r_{\rm th}-2 M)]}{\sqrt{c_2(b_{\rm th})}} \log{\left(\frac{b_{\rm th}}{b} -1\right)} \\
& -\frac{ 4 r_{\rm th}^2 [2M sa+b_{\rm th} (r_{\rm th}-2 M)] }{\sqrt{c_2(b_{\rm th})}} \log \left(\frac{\sqrt{-b_{\rm th} c_1'(b_{\rm th})}}{2 \sqrt{c_2(b_{\rm th})}}\right)
-\frac{4 M r_{\rm th}^2 (sa-b_{\rm th})}{\sqrt{c_2(b_{\rm th})}},\\
\end{split}
\ee
and the coefficients $\bar{a}$, $b_D$ can be read off as
\be
\bar{a} = \frac{2 r_{\rm th}^2 [2 M sa+b_{\rm th} (r_{\rm th}-2 M)]}{\sqrt{c_2(b_{\rm th})}} \label{bar_a<th}
\ee
and
\be \label{bar_b<th}
b_D =
-\bar{a} \log \left(\frac{-b_{\rm th} c_1'(b_{\rm th})}{4 c_2(b_{\rm th})}\right)
-\frac{4 M r_{\rm th}^2 (sa-b_{\rm th})}{\sqrt{c_2(b_{\rm th})}}.
\ee
The value of $b_R$ can be calculated from the integral of $f_R(z,b_{\rm th})$ through (\ref{b_R}).
Thus, the coefficient $\bar{b}$ is obtained from the sum of $b_D$ and $b_R$ in (\ref{bar_b}).
Then, the SDL deflection angle can be summarized as
\begin{equation} \label{hatalpha_as<}
\hat{\alpha}(b)\approx-\bar{a}\log{\left(\frac{b_c}{b}-1\right)}+\bar{b} \, .
\end{equation}

\begin{figure}[htp]
\begin{center}
    \includegraphics[width=15cm]{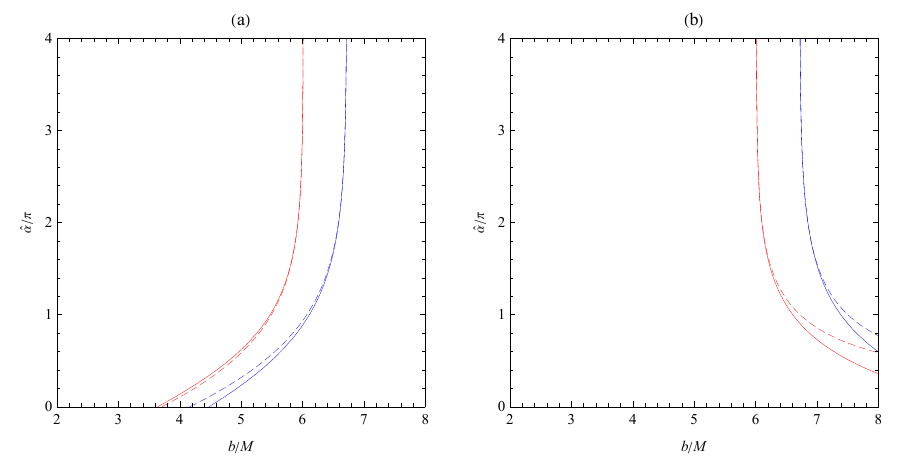}
    \caption{
    The SDL deflection angle $\hat{\alpha}(b)$ for $b<b_{\rm th}$ in (a) with (\ref{hatalpha_as<}),(\ref{bar_a<th}) and (\ref{bar_b<th}); for $b>b_{\rm th}$ in (b) with (\ref{hatalpha_as}), (\ref{bar_a>th}), and (\ref{bar_b>th}), for a direct orbit (red line) and a retrograde orbit (blue line) with $a/M=0.5$ and $\lambda=1.2$. The numerical results are displayed in dashed lines for comparison.}
    \label{def_angle_B}
\end{center}
\end{figure}

It is worth mentioning that the values of $\bar a$ and $b_D$ resulting from the approaches of $b>b_{\rm th}$ and $b<b_{\rm th}$ toward the double root in (\ref{bar_a>th}), (\ref{bar_b>th}), (\ref{bar_a<th}), (\ref{bar_b<th}) together with the respective $b_R$ are the same.
It is anticipated from Fig. \ref{effV}(b) that both approaches end up with the same case that the throat resides on the light sphere.
Figure \ref{barab_B} shows the values of $\bar a$ and $\bar b$, where $\bar a$ and $\vert \bar b \vert$ decrease in $\lambda$.
The deflection angle in the SDL is plotted in Fig. \ref{def_angle_B} showing good agreement with the numerical calculation of (\ref{deflection_angle}).

\subsection{The throat of a triple root: $r_{\rm th} =r_4 =r_3 \equiv {r}_{t}$ with the impact parameter $b_t$ at point C2}
\label{triple}
Another unique feature of the wormholes is that there exists a triple root of the radial potential $\tilde R(r)$ at the throat, namely $r_{\rm th} =r_4 =r_3 \equiv {r}_{t}$ with the parameters at point C2 in Fig. \ref{parameter_space}.
We will show that the divergence of the deflection angle is stronger than that of a logarithm.
The corresponding parameters are determined by ${r}_{sc}=r_{\rm th}$, where $r_{sc}$ obeys ${R}(r_{sc}) ={R}'(r_{sc}) =0$, leading to $\lambda=\lambda_t$ in (\ref{lambda0}).
The accompanying impact parameter $b(r_t)\equiv b_t$ can be found in (\ref{br0_k}).
The approaches from C1 to C2 for $b>b_{t}$ and from C3 to C2 for $ b<b_{t}$ will be discussed below.
\subsubsection{The case of $ b>b_{t}$ in the SDL and the approach from C1 to C2}
We consider the impact parameter $b$ in the region of $r_4 > r_{\rm th}$ with a value of $\lambda_t$, say at C1 and then changes the impact parameter from C1 toward C2.
Then, the turning point is set at $r_0=r_4$.
In terms of the variable $z$ in (\ref{z}), the deflection angle in (\ref{alpha}) then becomes (\ref{alpha_r0}) with the coefficients in (\ref{B}).
Note that when $r_0$ gets close to $r_{t}$ or $b \to {b}_{t}$, the turning point $r_0=r_4$ will simultaneously approach $r_3$ and $r_{\rm th}$.
Therefore, in this triple root case, the coefficients $c_1$ and $c_2$ in (\ref{c1_A>}) vanish when $r_0 \to r_t$,
\be
\begin{split}
c_1(r_0) &\sim \mathcal{O}(r_0-{r}_{t})^2 \to 0 \, , \\
c_2(r_0) &\sim \mathcal{O}(r_0-{r}_{t}) \to 0 \, , \\
\end{split}
\ee
whereas the other $c_n(r_{t})$ for $n>2$ is nonzero.

With that in mind, we introduce a new function, which gives the divergent part of the deflection angle after the integration over $z$ as $r_0$ approaches the triple root, to be
\be
f_D(z,r_0) \equiv \frac{2r_0^2 [r_0 b +2Mb(z-1) -2Msa(z-1)]}{\sqrt{c_1 z +c_2 z^2 +c_3 z^3}},
\ee
and then the integration of the remaining part $f_R(z,r_0) =f(z,r_0) -f_D(z,r_0)$ is finite.
Now we take the limit of $r_0=r_{t}$, where $c_1({r}_{t})$ and $c_2({r}_{t})$ vanish.
Then, the most significant contribution of the integral as $r_0 \to r_{t}$ comes from the $z \to 0$ region, where a natural lower limit can be set at $z\rightarrow \sqrt{{c_1}/{c_3}}$ or ${c_2}/{c_3}$, and ${c_1}/{c_2}= (r_0-{r}_{t})/{r_{t}} $ for $c_1 z\simeq c_2z^2\simeq c_3z^3$.
So, the divergent part can be obtained as follows
\be
\begin{split}
I_D(r_0)
\simeq & \int_0^1 \lim_{z\to 0} f_D(z,{r}_{t}) dz\\
= & \frac{2{r}_{t}^2 [{r}_{t} b_{t} -2Mb_{t} +2Msa] }{\sqrt{c_{3 t}}} \int_0^1 \frac{1}{z^{3/2}} dz +\mathcal{O}(z^{1/2}) \\
\sim & \left. \frac{4{r}_{t}^2 [{r}_{t} b_{t} -2Mb_{t} +2Msa] }{\sqrt{c_{3 t}}} z^{-1/2} \right|_{{z \to \frac{1}{r_{t}} (r_0-{r}_{t})} } + {\rm finite \, part}, \\
\end{split}
\ee
{with $c_{3 t} \equiv c_3(r_{t})$.}
In terms of the impact parameter given by (\ref{b_r_0}), the lower limit becomes
\be \label{z_b>bc}
z \to \sqrt{\frac{2}{b_{t}'' r_{t}^2}} (b-b_{t})^{1/2}\, ,
\ee
{with $b_{t}'' \equiv b''(r_{t})$. }
Then $I_D(b)$ can be approximated by
\be
I_D(b) \simeq \frac{4 {r}_{t}^{5/2} [2Msa +b_{t} (r_{t}-2M)]}{\sqrt{c_{3 t}}} \left( \frac{b_{t}''}{2 b_{t}} \right)^{1/4} \left( \frac{b}{b_{t}} -1 \right)^{-1/4}
+ {\rm finite \, part}.\\
\ee
As a result, the divergent part of the deflection angle has a power-law dependence of ${b}/{b_{t}} -1$ as $r_0 \to r_{t}$, which shows stronger divergence than logarithmic one due to a double root.
And the coefficients can be read off as
\be \label{bar_a_triple_b>bc}
\bar{a} = \frac{4 {r}_{t}^{5/2} [2Msa +b_{t} (r_{t}-2M)]}{\sqrt{c_{3 t}}} \left( \frac{b_{t}''}{2 b_{t}} \right)^{1/4}
\ee
and
\be \label{bar_b_triple_b>bc}
b_D = -\frac{4{r}_{t}^{2} [2Msa +b_{t} (r_{t}-2M)]}{\sqrt{{r}_{t}/M} \sqrt{c_{3 t}}},\\
\ee
where the finite part $b_D$ is obtained from the upper limit of the integral.
In the nonrotating limit, $a \to 0$, the coefficient $\bar{a}$ is reduced to a number $2.5558$ and $b_D =-4/3$ in Schwarzschild-like wormholes \cite{tsukamoto-2020}.

Moreover, the subleading term comes from the integration of $f_R(z,{r}_{sc})$, giving $b_R$ from (\ref{b_R}).
Again, the coefficient $\bar{b}$ can be calculated from the sum of $b_D$ and $b_R$ in (\ref{bar_b}).
To summarize, for the triple root, the deflection angle in the SDL is written in the form
\be\label{alpha_triple_root}
\hat{\alpha}(b) \approx \bar{a} \left(\frac{b}{b_{t}}-1\right)^{-1/4} + \bar{b} \\
\ee
with coefficients $\bar{a}>0$ and $\bar{b}$.
It is stressed again that the divergence of the deflection angle as $r_0$ approaches the triple root is stronger than that of the double root.

\begin{figure}[htp]
\begin{center}
    \includegraphics[width=15cm]{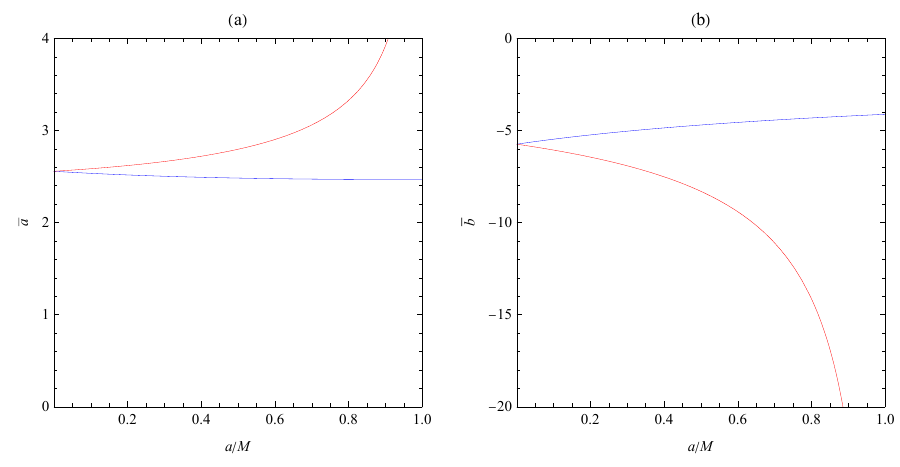}
    \caption{
    (a) the coefficients $\bar{a}$ as the functions of the spin $a$ with (\ref{bar_a_triple_b>bc}) and (\ref{bar_a_triple_b<bc}) for both $ b> b_{t}$ and $ b< b_{t}$, which has the same value;
    (b) the coefficients $\bar{b}$ as the functions of the spin $a$ mainly with (\ref{bar_b_triple_b>bc}) and (\ref{bar_b_triple_b<bc}) for both $ b> b_{t}$ and $ b< b_{t}$, which also has the same value, for a direct orbit (red line) with $\lambda=\lambda_t(r_{+c})$ and a retrograde orbit (blue line) with $\lambda =\lambda_t (r_{-c})$ in different wormholes.}
    \label{barab_C}
\end{center}
\end{figure}

\subsubsection{The case of $ b<b_{t}$ in the SDL and the approach from C3 to C2}
One can also approach the triple root from $b <b_{t}$ (from C3 to C2), where there is no turning point outside the throat and the throat is exactly on the light sphere seen in Figs. \ref{parameter_space} and \ref{effV}(c).
In terms of $z$ in (\ref{z_th}), the divergent part of the integral is given by
\be
f_D(z,b) \equiv \frac{2r_{t}^2 [r_{t} b +2Mb(z-1) -2Msa(z-1)]}{\sqrt{c_1 z +c_2 z^2 +c_3 z^3}},
\ee
leading to the deflection angle in (\ref{alpha_th}) with the coefficients in (\ref{B_th}) where $c_1$ and $c_2$ in (\ref{c_B<}) become
\be
\begin{split}
c_1(b) &\sim \mathcal{O}(b_{t}-b) \to 0 \, , \\
c_2(b) &\sim \mathcal{O}(b_{t}-b) \to 0 \, .\\
\end{split}
\ee
We then obtain the divergent part by taking the limit of $z\to 0$ and setting $b=b_{t}$ where $c_1(b_{t})=c_2(b_{t})=0$ in the integrand as follows
\be
\begin{split}
I_D(b) \simeq & \int_0^1 \lim_{z\to 0} f_D(z,{b}_{t}) dz\\
       = & \frac{2{r}_{t}^2 [{r}_{t} b_{t} -2Mb_{t} +2Msa] }{\sqrt{c_{3}({b}_{t})}} \int_0^1 \frac{1}{z^{3/2}} dz +\mathcal{O}(z^{1/2}) \\
       \sim & \left. \frac{4{r}_{t}^2 [{r}_{t} b_{t} -2Mb_{t} +2Msa] }{\sqrt{c_{3}({b}_{t})}} z^{-1/2} \right|_{z \to \sqrt{\frac{c_1}{c_3}}}
       + {\rm finite \, part} ,\\
\end{split}
\ee
where the lower limit can be set at
\be
z \to \sqrt{\frac{c_1}{c_3}} = \sqrt{\frac{ -c_1'(b_{t})}{c_3(b_{t})}} (b_{t}-b)^{1/2},
\ee
which is the same as in (\ref{z_b>bc}) with the different expression.
In the end, the divergent part becomes
\be
I_D(b) \simeq \frac{4{r}_{sc}^2 [2Msa +b_{t} (r_{t}-2M)] }{\sqrt{c_{3}(b_{t})}} \left( \frac{c_3(b_{t})}{-b_{t} c_1'(b_{t})} \right)^{1/4} \left( \frac{b_{t}}{b}-1 \right)^{-1/4} + {\rm finite \, part}, \\
\ee
with the coefficients
\be \label{bar_a_triple_b<bc}
\bar{a} = \frac{4{r}_{t}^{2} [2Msa +b_{t} (r_{t}-2M)]}{\sqrt{c_{3}({r}_{t})}} \left( \frac{c_3(b_{t})}{-b_{t} c_1'(b_{t})} \right)^{1/4}
\ee
and
\be \label{bar_b_triple_b<bc}
b_D = -\frac{4{r}_{t}^{2} [2Msa +b_{t} (r_{t}-2M)]}{\sqrt{{r}_{t}/M} \sqrt{c_{3}({b}_{t})}} .\\
\ee
Again, the finite part $b_D$ can be obtained from the upper limit of the integral.
To summarize, for the triple root, the SDL deflection angle is in the form
\be\label{alpha_triple_root<}
\hat{\alpha}(b) \approx \bar{a} \left(\frac{b_{t}}{b}-1\right)^{-1/4} + \bar{b}. \\
\ee

\begin{figure}[htp]
\begin{center}
    \includegraphics[width=15cm]{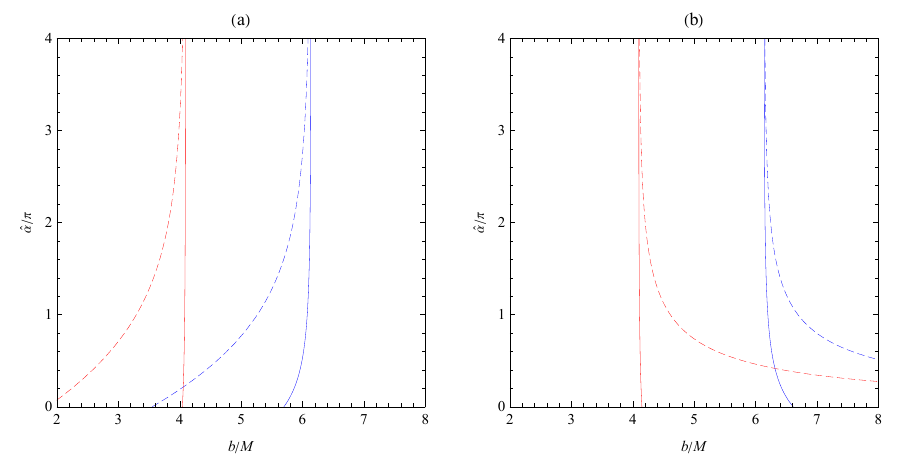}
    \caption{
    The SDL deflection angle $\hat{\alpha}(b)$ for $b<b_{t}$ in (a) with (\ref{alpha_triple_root<}); for $b>b_{t}$ in (b) with (\ref{alpha_triple_root}), for a direct orbit (red line) with $a/M=0.5$, $\lambda=\lambda_t(r_{+c})$ and a retrograde orbit (blue line) with $a/M=0.5$, $\lambda =\lambda_t (r_{-c})$ in different wormholes.
    The numerical solutions are shown in dashed lines for comparison.}
    \label{def_angle_C}
\end{center}
\end{figure}

Based on the effective potential $w_\text{eff}(l)$ in Fig. \ref{effV}(c), we expect that $\bar a$ and $\bar b$ from $b <b_{t}$ are the same as their counterparts from $b >b_{t}$.
Figure \ref{barab_C} shows the values of $\bar a$ and $\bar b$ with the deflection angle of the power-law divergence at the triple root from the approaches of $b>b_{t}$ and $b<b_{t}$ in Fig. \ref{def_angle_C}.
A good agreement between the numerical calculation of (\ref{deflection_angle}) and the above analytical approximate solutions in the SDL is shown.

\subsection{The throat outside the double root: $r_{\rm th} >r_4=r_3\equiv r_{sc}$}
\label{throat_outside}
For completeness, let us also consider the case $r_4=r_3 < r_{\rm th}$ with the parameter at D in Fig. \ref{parameter_space}, although the light rays will not reach the double root and only meet the throat.
We start by choosing the impact parameter $b$ in the region of $r_3< r_4 < r_{\rm th}$ in the light orange region of Fig. \ref{parameter_space}.
In particular, when $r_{\rm th} > r_4$ of a unique scenario for wormholes, the light rays can pass through the throat to another spacetime without reaching the roots of $r_3$ and $r_4$, which can be visualized in Fig. 6 in \cite{taylor-2014}.
We introduce the variable
\be
u \equiv \frac{r_{\rm th}}{r},
\ee
and then the integration in (\ref{alpha}) can be rewritten as
\be
\hat{\alpha}+\pi = 2 \int_{0}^{1} f_T(u,b) du,
\ee
where the integrated function is
\be
f_T(u,b) =\frac{\frac{2M sa}{r_{\rm th}^2}u +\frac{b}{r_{\rm th}} \left(1-\frac{2 M}{r_{\rm th}}u\right)}{\sqrt{
\left(1 -\frac{2 M }{r_{\rm th}}u +\frac{a^2 }{r_{\rm th}^2}u^2 \right)
\left(1 +\frac{a^2-b^2}{r_{\rm th}^2}u^2 +\frac{2 M (sa-b)^2}{r_{\rm th}^3}u^3 \right)
\left(1 -u \right)
\left(1 -\frac{a^2 }{r_{\rm th}^2} u \right)}}.
\ee
Also, we rewrite the spin as the unit spin $a_{\bullet} \equiv a/M$, and the impact parameter as a unit impact parameter $b_\bullet \equiv b/M$.
Then to see the significantly different lensing effect by wormholes than black holes considers that the throat of the wormhole $r_{\rm th}$ is much larger than $r_+$ with the scale $M$, which is a horizon scale of Kerr black holes when $\lambda=0$.
So, we expand the integrated function $f_T(u,b_\bullet)$ in terms of small $M/r_{\rm th}$ as
\be
f_T(u,b_\bullet)
= \frac{b_\bullet }{\sqrt{1-u} } \left( \frac{M}{r_{\rm th}} \right)
-\frac{ (2 sa_\bullet -b_\bullet) u }{\sqrt{1-u}} \left( \frac{M}{r_{\rm th}} \right)^2
+\mathcal{O}\left( \frac{M}{r_{\rm th}} \right)^3 .
\ee

Ultimately, by integrating the expansion order by order, we obtain the deflection angle as
\be \label{b_th}
\hat{\alpha}(b_\bullet) = -\pi
+ 4b_\bullet \left( \frac{M}{r_{\rm th}} \right)
+ \frac{8}{3} (2 sa_\bullet -b_\bullet) \left( \frac{M}{r_{\rm th}} \right)^2
+\mathcal{O}\left( \frac{M}{r_{\rm th}} \right)^3 \, .
\ee
Toward the limit of $r_4 =r_3=r_{sc}$, the impact parameter is $b_\bullet = b_{sc}/M$.
The expression of the deflection angle looks as if the light rays are reflected back to the light source because of $-\pi$.
In fact, we should interpret that light rays are scattered to another spacetime by the throat.
The large deflection angle can be visualized in the plot in the middle right of Fig. 6 in \cite{taylor-2014}.

\section{Deflection angle in the weak deflection limit}
\label{secIV}
In this section, we are interested in the deflection angle in the other limit, which means that the closest distance $r_0$ is much larger than the scale of the throat $r_{\rm th}$ and $r_{+}$, which is of order $M \approx M^{(\text{ADM})}_{\text{WH}}$ for $\lambda$ of order one.
The corresponding parameters for $r_4 \gg r_{\rm th}$ are {in the light green and yellow regions} of Fig. \ref{parameter_space}.
In the weak field limit (WDL), the deflection angle is expectedly small.
Therefore, we choose the ratio $M/r_0 \ll 1$ as a small parameter, and the deflection angle in the WDL can be expressed order by order with respect to this small ratio.

The closest distance $r_0$ is a turning point of the orbit, that is, the outermost single root $r_4$.
The light rays will travel far from the throat during their journey.
Now we first introduce the variable $u$
\be
u \equiv \frac{r_0}{r},
\ee
and then the deflection angle can be rewritten as
\be
\hat{\alpha}+\pi = 2 \int_{0}^{1} f_W(u,r_0) du,
\ee
where the integrated function is given by
\be
f_W(u,r_0) =\frac{\frac{2M sa }{r_0^2}u +\frac{b}{r_0} \left(1-\frac{2 M}{r_0}u\right)}{\sqrt{\left(1 -\frac{2 M }{r_0}u +\frac{a^2 }{r_0^2}u^2 \right)
\left(1 +\frac{a^2-b^2}{r_0^2}u^2 +\frac{2 M (sa-b)^2}{r_0^3}u^3 \right) \left(1 -\frac{2 M (1+\lambda ^2)}{r_0}u +\frac{a^2}{r_0^2}u^2 \right) }}.
\ee
The impact parameter $b(r_0)$ for the turning point $r_4$ is obtained from (\ref{br0_k}).
In terms of the unit spin $a_{\bullet}$ of the wormhole, the expression of $f_W(u,r_0)$ can be written as a function of the small ratio $M/r_0$.
We then expand the function $f_W(u,r_0)$ in terms of $M/r_0$ as
\be
\begin{split}
f_W(u&, r_0) \\
=& \frac{1}{\sqrt{1-u} \sqrt{1+u}}
+\frac{ \left(\lambda ^2+1\right) u^2+\left(\lambda ^2+1\right) u+1 }{\sqrt{1-u} (1+u)^{3/2}} \left( \frac{M}{r_0} \right) \\
& \scalebox{1.2}{$ -\frac{ a_{\bullet}^2 (1+u)^2 \left(2 u^2-1\right)+4 sa_{\bullet} (1+u)-3 \left(\lambda ^2+1\right)^2 u^4-6 \left(\lambda ^2+1\right)^2 u^3-\left(3 \lambda ^4+8 \lambda ^2+9\right) u^2-2 \left(\lambda ^2+3\right) u-3 }{2 \sqrt{1-u} (1+u)^{5/2}} \left( \frac{M}{r_0} \right)^2 $} \\
&+\mathcal{O}\left( \frac{M}{r_0} \right)^3.
\end{split}
\ee
The integration of $f_W(u,r_0)$ can be carried out order by order.
Up to the order of $(M/r_0)^2$, the deflection angle is given by
\be
\hat{\alpha}(r_0) = 2 \left(\lambda ^2+2\right) \left( \frac{M}{r_0} \right)
+ \left[-4 sa_{\bullet}-2 \left(\lambda ^2+2\right)+\frac{\pi}{4} \left(3 \lambda ^4+10 \lambda ^2+15\right)\right] \left( \frac{M}{r_0} \right)^2
+\mathcal{O}\left( \frac{M}{r_0} \right)^3 .
\ee
Finally, let us replace $r_0$ by $b$ using (\ref{r0b}), so the deflection angle in the WDL becomes
\be
\hat{\alpha}(b) = 2 \left(\lambda ^2+2\right) \left( \frac{M}{b} \right)
+\left[ \frac{\pi}{4} \left(3 \lambda ^4+10 \lambda ^2+15\right)-4 sa_{\bullet} \right] \left( \frac{M}{b} \right)^2
+\mathcal{O}\left( \frac{M}{b} \right)^3 \, .
\ee
%
Then, two special limits are considered.
In the case of a Kerr black hole by setting the parameter $\lambda =0$ or the throat $r_{\rm th \bullet}=r_+/M$, the deflection angle is reduced to
\be
\hat{\alpha}(b) = \frac{4M}{b}
+\left[ \frac{15 \pi}{4} -4 sa_{\bullet} \right] \left( \frac{M}{b} \right)^2
+\mathcal{O}\left( \frac{M}{b} \right)^3 \, ,
\ee
consistent with \cite{iyer-2009}.
When turning off the spin $a=0$, which reduces to Schwarzschild-like wormholes, the deflection angle is then
\be
\hat{\alpha}(b) = 2 \left(\lambda ^2+2\right) \left( \frac{M}{b} \right)
+\left[ \frac{\pi}{4} \left(3 \lambda ^4+10 \lambda ^2+15\right) \right] \left( \frac{M}{b} \right)^2
+\mathcal{O}\left( \frac{M}{b} \right)^3,
\ee
also consistent with the leading order result in \cite{tsukamoto-2020}.

More importantly, we also find that the spin effect of the Kerr-like wormhole will not appear until the order $(M/b)^2$, but the throat effect always exists in each orders.
In other words, the farther away the light ray is from the wormhole, the more the deflection angle looks like a Schwarzschild-like wormhole.
Our results are different from \cite{ovgun-2018} in the subleading order.

\section{Gravitational lens and relativistic images}
\label{secV}
As for gravitational lens and its relativistic images, we consider the planar light rays with the lens diagram in Fig. \ref{fig:arch_02} as an illustration
 when the observers and the light sources are in the same spacetime.
The distances of the lens (wormhole) and the light source from the observer are denoted by $d_L$ and $d_S$ whereas $d_{LS}$ represents the distance between the lens and the source.
A reference optical axis is defined by the line connecting the observer and the lens so that the angular positions of the source and the image denoted by $\beta$ and $\theta$, respectively, are measured from the optical axis. The lens equation is given by \cite{virbhadra-2000}
\begin{equation} \label{leneq}
\tan s\beta=\tan\theta- \frac{d_{LS}}{d_S}[\tan\theta+\tan(\hat{\alpha}-\theta)]\;,
\end{equation}
where $\hat{\alpha}$ is the deflection angle of light rays obtained from (\ref{deflection_angle}) that can also be expressed in terms of the impact parameter $b$.
The deflection angle $\hat \alpha$ can be approximately by {(\ref{hatalpha_as}) and (\ref{hatalpha_as_below}) for the double root or (\ref{hatalpha_triple}) and (\ref{hatalpha_triple_below}) } for the triple root.
In the SDL of our interest, the light rays wind around the wormhole $n$ times.
The angle in the lens equation is within $2\pi$ to be achieved from the deflection angle $\hat \alpha$ subtracting $ 2n\pi$.

Together with the relation between the impact parameter $b$ and the angular position of the image given by
\begin{equation} \label{b_theta}
b=d_L \sin\theta
\end{equation}
in Fig. \ref{fig:arch_02}, we can solve the lens equation (\ref{leneq}) with a given angular position of the source $\beta$ for the angular position of the observed image $\theta$ {by following the approach of \cite{bozza-2002, hsieh-2021A}.}
{In the SDL, since the impact parameter is of order $b_c$, giving $b \sim b_c \ll d_L$, Eq. (\ref{b_theta}) can be further approximated by $ b \approx d_L \theta$.
We have shown that the SDL deflection angle will increase quickly to infinity when the impact parameter approaches the critical value.
Now we consider the deflection angle $\hat{\alpha} =2 n \pi +\delta \hat{\alpha}$, where $\delta \hat{\alpha}\ll 2\pi$ is a small angle.}
In this case, $\beta$ is expectedly small.
As a result, the lens equation (\ref{leneq}) can be further simplified by
\begin{equation}\label{leneq_app}
s\beta\simeq \theta-\frac{d_{LS}}{d_{S}}[\hat{\alpha}(\theta)-2n\pi] \,.
\end{equation}
{Then, we expand the deflection angle $\hat{\alpha}(\theta)$ around $\theta =\theta_{sn}^0$ as
\be
\hat{\alpha}(\theta) =\hat{\alpha}(\theta_{sn}^0) + \left.\hat{\alpha}'(\theta)\right|_{\theta =\theta_{sn}^0} (\theta-\theta_{sn}^0)+\mathcal{O}(\theta-\theta_{s n}^0)^2\, ,
\ee
where $\hat{\alpha}(\theta_{sn}^0)=2n\pi$ serves as the zeroth order solution.}
Plugging the expansion into the approximate lens equation (\ref{leneq_app}), to the linear order $(\theta-\theta_{sn}^0)$, it becomes
\be
s\beta\simeq\theta_{sn}^0 +\left(1 -\frac{d_{LS}}{d_S} \hat{\alpha}'(\theta_{sn}^0) \right) (\theta-\theta_{sn}^0) \, .
\ee
Solving for $\theta$, the angular position of the image is found to be \cite{bozza-2002}
\be \label{theta_sn}
\theta_{s n} \simeq \theta_{sn}^0 + \left(1 -\frac{d_{LS}}{d_S} \hat{\alpha}'(\theta_{sn}^0) \right)^{-1} (s \beta -\theta_{sn}^0) \;.
\ee
The deflection angle plays a vital role to determine the angular positions of the relativistic images.
Based on the effective potentials in Figs. \ref{effW} and \ref{effV}, the light rays can also possibly pass through the throat and travel from one spacetime to another to be observed.
According to \cite{shaikh-2019}, due to the symmetry of defining the proper radial distance, $l \to -l$, the light rays from the sources can be mapped into the spacetime of the observers so that the lens equation in (\ref{leneq_app}) is assumed to be the same.

\begin{table}
\caption{\label{tab:table-I_s}
    \baselineskip=18pt
    Relativistic images for the observer and the light source, which are in a single spacetime, are in the same side of the source with the angular position $\beta=10$ ($\mu$as), where the light rays are along direct orbits seen in Fig. \ref{fig:arch_02} with $\lambda=0.1<\lambda_t$ (double root) from (\ref{theta_sn_d}) and $b_{+c}$ from (\ref{bsc_k}), with $\lambda=\lambda_t$ (triple root) from (\ref{image_s_triple}) and $b_{+c}$ from (\ref{bsc_k}), and with $\lambda=1.2>\lambda_t$ (double root) from (\ref{theta_sn_d}) and $b_{\rm th}$ from (\ref{br0_k}) for the wormhole spin $a/M$.}
\begin{tabular}{ccccccc}
\hline
\hline
$a/{M}$~  & $\lambda$~ &$\theta_{+1}$ ($\mu$as) &$\hat{\alpha}$ &$b/M$  &$\theta_{+\infty}$ ($\mu$as) & $\Delta \theta_{+}$ ($\mu$as)\\
\hline
$10^{-3}$ & $0.1$     &26.4250 & $2\pi+32.8104$ ($\mu$as) & $5.2010$    &26.3900 &0.0350\\
$ $       & $0.7067$  &26.4437 & $2\pi+32.9329$ ($\mu$as) & $5.2047$    &26.3900 &0.0537\\
$ $       & $1.2$     &32.8007 & $2\pi+45.6084$ ($\mu$as) & $6.4559$    &32.2709 &0.5298\\
\hline
$0.5$     & $0.1$     &20.9372 & $2\pi+21.8851$ ($\mu$as) & $4.1209$    &20.8120 &0.1252\\
$ $       & $0.4763$  &20.8400 & $2\pi+21.8037$ ($\mu$as) & $4.1018$    &20.8120 &0.0280\\
$ $       & $1.2$     &31.0020 & $2\pi+42.0041$ ($\mu$as) & $6.1019$    &30.5472 &0.4548\\
\hline
$0.9$     & $0.1$     &15.0431 & $2\pi+10.0870$ ($\mu$as) & $2.9608$     &14.4517 &0.5914\\
$ $       & $0.1972$  &14.4571 & $2\pi+ 9.3220$ ($\mu$as) & $2.8455$     &14.4517 &0.0055\\
$ $       & $1.2$     &29.5767 & $2\pi+39.1513$ ($\mu$as) & $5.8214$     &29.1486 &0.4281\\
\hline
\hline
\end{tabular}
\end{table}

\subsection{For the double root: $\theta_{s n} > \theta_{s\infty}$ for $ b>b_c$; $\theta_{s n} < \theta_{s\infty}$ for $ b<b_c$}
We start from the double root with the critical impact parameter $b_c$, which is $r_4=r_3=r_{sc}$ with $b_{sc}$ or $r_4=r_{\rm th}$ with $b_{\rm th}$, and is approached from $b > b_c$.
The SDL deflection angle has the logarithmic form as in (\ref{hatalpha_as}) in \cite{hsieh-2021A}, giving that the angular position of the relativistic image (\ref{theta_sn}) is expressed as,
\begin{equation} \label{theta_sn_d}
\theta_{s n} \simeq \theta_{sn}^0+\frac{ e^{(\bar{b}-2n\pi)/\bar{a}}}{ \bar a} \frac{ b_{c} d_S}{d_{LS}d_L } (s \beta -\theta_{sn}^0) \;
\end{equation}
for $b_{c}/d_L \ll 1 $.
{The zeroth order solution is obtained from solving $\hat{\alpha}(\theta_{sn}^0)=2n\pi$ for $n=1,2,\cdots$, using (\ref{hatalpha_as}) and $b\approx d_L \theta$, and is given by}
\begin{equation} \label{theta0}
\theta_{sn}^0=\frac{ b_{c}}{d_L}\left( 1+e^{\frac{\bar{b}-2n\pi}{\bar{a}}} \right)\, .
\end{equation}
{Because in (\ref{hatalpha_as}) the impact parameter $b$ is larger than its critical value $b_c$, the angular position $\theta_{sn}$ lies outside $\theta_{s \infty}$ with the value $\theta_{s \infty}={ b_{c}}/{d_L}$ in terms of the critical impact parameter.}
The value of $\theta_{sn}$ decreases as the number of laps $n$ increases, and approaches $\theta_{s \infty}$ as $n \to \infty$.
Here, the coefficients $\bar{a}$, $\bar{b}$, and the critical impact parameters $b_{sc}$ or $b_{\rm th}$ depend on the parameters of the Kerr-like wormhole, leading to the modification of the coefficients as compared with Kerr black holes.

There is another way to approach the double roots $r_3 =r_4 =r_{sc}$ and $r_4 =r_{\rm th}$, that is, $b \to b_{c}$ from $b <b_{c}$, where the observers and the light sources are in different spacetimes.
{The corresponding SDL deflection angle is expressed in (\ref{hatalpha_as_below}).}
The zeroth order $\theta_{sn}^0$ solution now becomes
\begin{equation} \label{theta_infty}
\theta_{sn}^0=\frac{ b_{c}}{d_L}\left( 1+e^{\frac{\bar{b}-2n\pi}{\bar{a}}} \right)^{-1}.
\end{equation}
%
%
The angle of the relativistic image in (\ref{theta_sn}) with the input of the SDL deflection angle (\ref{hatalpha_as_below}) is obtained as
\begin{equation}
\theta_{s n} \simeq \theta_{sn}^0 -\frac{ e^{(\bar{b}-2n\pi)/\bar{a}}}{ \bar{a} [1+e^{(\bar{b}-2n\pi)/\bar{a}} ]^2} \frac{ b_{c} d_S}{d_{LS}d_L } (s \beta -\theta_{sn}^0) \;.
\label{image_a_double}
\end{equation}
{Due to the fact that in (\ref{hatalpha_as_below}) the impact parameter $b$ is smaller than $b_c$, the angular position $\theta_{sn}$ lies inside $\theta_{s \infty}$ given by $\theta_{s \infty}={ b_{c}}/{d_L}$.}
The value of $\theta_{sn}$ then increases when $n$ increases in the spacetime of the observers, and will also approach $\theta_{s \infty}$ when $n \to \infty$.
The scenario of $\theta_{s n} < \theta_{s\infty}$ for $ b<b_c$ can only occur in wormhole where the light source and the observer are in different spacetimes.
%

\begin{table}
\caption{\label{tab:table-II_s}
    \baselineskip=18pt
    Relativistic images for the observer and the light source, which are in a single spacetime, are in the opposite side of the source with the angular position $\beta=10$ ($\mu$as), where the light rays are in retrograde orbits seen in Fig. \ref{fig:arch_02} with $\lambda=0.1<\lambda_t$ (double root) from (\ref{theta_sn_d}) and $b_{-c}$ from (\ref{bsc_k}), with $\lambda=\lambda_t$ (triple root) from (\ref{image_s_triple}) and $b_{-c}$ from (\ref{bsc_k}), and with $\lambda=1.2>\lambda_t$ (double root) from (\ref{theta_sn_d}) and $b_{\rm th}$ from (\ref{br0_k}) for the wormhole spin $a/M$.}
\begin{tabular}{ccccccc}
\hline
\hline
$a/{M}$~  & $\lambda$~ &$\theta_{-1}$ ($\mu$as) &$\hat{\alpha}$ &$b/M$  &$\theta_{-\infty}$ ($\mu$as) & $\Delta \theta_- $ ($\mu$as)\\
\hline
$10^{-3}$ & $0.1$     &26.4451 & $2\pi+72.8586$ ($\mu$as) & $5.2050$    &26.4103 &0.0348\\
$ $       & $0.7075$  &26.4641 & $2\pi+72.9200$ ($\mu$as) & $5.2087$    &26.4103 &0.0349\\
$ $       & $1.2$     &32.8081 & $2\pi+85.6189$ ($\mu$as) & $6.4574$    &32.2779 &0.5302\\
\hline
$0.5$     & $0.1$     &31.2000 & $2\pi+82.4656$ ($\mu$as) & $6.1409$    &31.1862 &0.0138\\
$ $       & $0.8952$  &31.2673 & $2\pi+82.4924$ ($\mu$as) & $6.1541$    &31.1862 &0.0811\\
$ $       & $1.2$     &34.8104 & $2\pi+89.6243$ ($\mu$as) & $6.8515$    &34.1400 &0.6704\\
\hline
$0.9$     & $0.1$     &34.7206 & $2\pi+89.2466$ ($\mu$as) & $6.8338$    &34.7131 &0.0075\\
$ $       & $1.0289$  &34.8185 & $2\pi+89.5763$ ($\mu$as) & $6.8531$    &34.7131 &0.1054\\
$ $       & $1.2$     &36.7321 & $2\pi+93.4704$ ($\mu$as) & $7.2297$    &35.9030 &0.8291\\
\hline
\hline
\end{tabular}
\end{table}

\subsection{For the triple root: $\theta_{s n} > \theta_{s\infty}$ for $ b>b_t$; $\theta_{s n} < \theta_{s\infty}$ for $ b<b_t$}

\begin{table}
\caption{\label{tab:table-I_d}
    \baselineskip=18pt
    Relativistic images for the observer and the light source, which are in different spacetimes, are in the same side of the source by mapping $l \to -l$ to the same spacetime of the observer with the angular position $\beta=10$ ($\mu$as), where the light rays are along direct orbits seen in Fig. \ref{fig:arch_02} with $\lambda=0.1<\lambda_t$ (double root) from (\ref{image_a_double}) and $b_{+c}$ from (\ref{bsc_k}), with $\lambda=\lambda_t$ (triple root) from (\ref{image_a_triple}) and $b_{+c}$ from (\ref{bsc_k}), and with $\lambda=1.2>\lambda_t$ (double root) from (\ref{image_a_double}) and $b_{\rm th}$ from (\ref{br0_k}) for the wormhole spin $a/M$.}
\begin{tabular}{ccccccc}
\hline
\hline
$a/{M}$~  & $\lambda$~ &$\theta_{+1}$ ($\mu$as) &$\hat{\alpha}$ &$b/M$  &$\theta_{+\infty}$ ($\mu$as) & $\Delta \theta_{+}$ ($\mu$as)\\
\hline
$10^{-3}$ & $0.1$     &25.3624 & $2\pi+30.7245$ ($\mu$as) & $4.9919$    &26.3900 &-1.0276\\
$ $       & $0.7067$  &26.3364 & $2\pi+32.6869$ ($\mu$as) & $5.1836$    &26.3900 &-0.0536\\
$ $       & $1.2$     &31.7496 & $2\pi+43.4953$ ($\mu$as) & $6.2491$    &32.2709 &-0.5212\\
\hline
$0.5$     & $0.1$     &18.1976 & $2\pi+16.3942$ ($\mu$as) & $3.5817$    &20.8120 &-2.6144\\
$ $       & $0.4763$  &20.7840 & $2\pi+21.5553$ ($\mu$as) & $4.0908$    &20.8120 &-0.0280\\
$ $       & $1.2$     &30.0990 & $2\pi+40.1985$ ($\mu$as) & $5.9242$    &30.5472 &-0.4482\\
\hline
$0.9$     & $0.1$     &12.1309  & $2\pi+4.2612$ ($\mu$as) & $2.3876$    &14.4517 &-2.3208\\
$ $       & $0.1972$  &14.4462  & $2\pi+9.6715$ ($\mu$as) & $2.8434$    &14.4517 &-0.0055\\
$ $       & $1.2$     &28.7267  & $2\pi+37.4494$ ($\mu$as) & $5.6541$    &29.1486 &-0.4219\\
\hline
\hline
\end{tabular}
\end{table}

For the triple root with the critical impact parameter $b_t$, the SDL deflection angle has the power-law divergence in (\ref{hatalpha_triple}) as the triple root is approached from both $b > b_t$ and $b < b_t$, which is a unique scenario for wormhole.
Let us now consider the approach of a triple root from $b>b_t$.
Again, the zeroth order solution $\theta_{sn}^0$ is determined by $\hat{\alpha}(\theta_{sn}^0)=2n\pi$.
With (\ref{hatalpha_triple}) and $b\approx d_L \theta$, the solution is found to be
\be \label{theta0_triple_root}
\theta_{sn}^0 =\frac{ b_{t}}{d_L} \left[ 1 +\left( \frac{2n\pi -\bar{b}}{\bar{a}} \right)^{-4} \right]
\ee
for $n=1,2,\cdots$.
{Substituting the SDL deflection angle (\ref{hatalpha_triple}) into (\ref{theta_sn}), the angular position of the image then becomes}
\be \label{image_s_triple}
\theta_{s n}\simeq\theta_{sn}^0 +\frac{4 {b}_{t} d_S \bar{a}^4}{ d_L d_{LS} (2n\pi -\bar{b})^5} (s \beta -\theta_{sn}^0) \;
\ee
for ${b}_{t}/d_L \ll 1 $.
The angular position $\theta_{sn}$ is outside $\theta_{s \infty}$ given by $n \to \infty$, and decreases when $n$ increases.

One can approach the triple root from $b <b_{t}$ where the observers see the light rays come from another spacetime with the SDL deflection angle in (\ref{hatalpha_triple_below}), giving the angular position in (\ref{theta_sn}) as
\be
\theta_{s n}\simeq\theta_{sn}^0
-\frac{4 {b}_{t} d_S \bar{a}^4}{ d_L d_{LS} (2n\pi -\bar{b})^5} \left[ \left( \frac{2n\pi -\bar{b}}{\bar{a}} \right)^{-4} +1 \right]^{-2} (s \beta -\theta_{sn}^0)
\label{image_a_triple}
\ee
with
\be \label{theta_infty_t}
\theta_{sn}^0 =\frac{ b_{t}}{d_L} \left[ 1 +\left( \frac{2n\pi -\bar{b}}{\bar{a}} \right)^{-4} \right]^{-1}\, .
\ee
Again, the angular position $\theta_{sn}$ is inside $\theta_{s \infty}$ and increases as $n$ increases.

We now compute the angular positions of the relativistic images of the sources for $n=1$ ($\theta_{\pm 1}$) by the Kerr-like wormholes with mass $M^{(\text{ADM})}_{\text{WH}}\approx M$, where $M=4.1\times 10^6 M_{\odot}$ and the distance $d_L=26000\;{\rm ly}$ of the scale of the supermassive black hole Sagittarius A* at the center of our Galaxy.
We also take the ratio to be $d_{LS}/d_S=1/2$.
We consider in Table \ref{tab:table-I_s} (\ref{tab:table-II_s}), the observers and the light sources are in a single spacetime, whereas in Table \ref{tab:table-I_d} (\ref{tab:table-II_d}), the light rays are from another spacetime to be observed.
We also consider both the image and the source to be in the same (opposite) side of the optical axis, where the light rays travel along the direct (retrograde) orbit.
The angular positions of the relativistic images are computed by (\ref{theta_sn_d}), (\ref{image_a_double}), (\ref{image_s_triple}), and (\ref{image_a_triple}).
In the case of $b_{c} \ll d_L$, $\theta_{sn}$ is not sensitive to $\beta$ but is mainly determined by $\theta_{sn}^0$ in (\ref{theta0}), (\ref{theta_infty}), (\ref{theta0_triple_root}), and (\ref{theta_infty_t}).
Apart from the fact that the images are distributed with respect to the reference optical axis
in an asymmetric way, another common trend is that the angular position decreases (increases) with the wormhole spin $a$ for the direct (retrograde) orbit.
Notice that since $b \approx d_L \theta$ for $b>b_c$ (for $ b<b_c$), $\theta$ decreases (increases) from $\theta_{\pm 1}$ to $\theta_{\pm\infty}$.
\begin{table}
\caption{\label{tab:table-II_d}
    \baselineskip=18pt
    Relativistic images for the observer and the light source, which are in different spacetimes, are in the opposite side of the source by mapping $l \to -l$ to the same spacetime of the observers with the angular position $\beta=10$ ($\mu$as), where the light rays are in retrograde orbits seen in Fig. \ref{fig:arch_02} with $\lambda=0.1<\lambda_t$ (double root) from (\ref{image_a_double}) and $b_{-c}$ from (\ref{bsc_k}), with $\lambda=\lambda_t$ (triple root) from (\ref{image_a_triple}) and $b_{-c}$ from (\ref{bsc_k}), and with $\lambda=1.2>\lambda_t$ (double root) from (\ref{image_a_double}) and $b_{\rm th}$ from (\ref{br0_k}) for the wormhole spin $a/M$.}
\begin{tabular}{ccccccc}
\hline
\hline
$a/{M}$~  & $\lambda$~ &$\theta_{-1}$ ($\mu$as) &$\hat{\alpha}$ &$b/M$  &$\theta_{-\infty}$ ($\mu$as) & $\Delta \theta_- $ ($\mu$as)\\
\hline
$10^{-3}$ & $0.1$     &25.3870 & $2\pi+70.7760$ ($\mu$as) & $4.9967$    &26.4103 &-1.0234\\
$ $       & $0.7075$  &26.3567 & $2\pi+72.7245$ ($\mu$as) & $5.1876$    &26.4103 &-0.0537\\
$ $       & $1.2$     &31.7563 & $2\pi+83.5036$ ($\mu$as) & $6.2504$    &32.2779 &-0.5216\\
\hline
$0.5$     & $0.1$     &30.9951 & $2\pi+81.9871$ ($\mu$as) & $6.1005$    &31.1862 &-0.1911\\
$ $       & $0.8952$  &31.1053 & $2\pi+82.2747$ ($\mu$as) & $6.1222$    &31.1862 &-0.0809\\
$ $       & $1.2$     &33.4825 & $2\pi+86.9603$ ($\mu$as) & $6.5901$    &34.1400 &-0.6575\\
\hline
$0.9$     & $0.1$     &34.7087 & $2\pi+89.0440$ ($\mu$as) & $6.8315$    &34.7131 &-0.0043\\
$ $       & $1.0289$  &34.6079 & $2\pi+89.2774$ ($\mu$as) & $6.8116$    &34.7131 &-0.1051\\
$ $       & $1.2$     &35.0926 & $2\pi+90.1820$ ($\mu$as) & $6.9070$    &35.9030 &-0.8104\\
\hline
\hline
\end{tabular}
\end{table}
For the double root, with a given spin $a$, as $\lambda$ increases but is still less than $\lambda_t$, we observe a significant increase in the angular position $\theta_{+1}$ given by a direct orbit in Table \ref{tab:table-I_s}, especially for the near extremal Kerr-like wormhole $a \to M$.
In contrast, the increase in the angular position $\theta_{-1}$ given by a retrograde orbit in Table \ref{tab:table-II_s} is less noticeable.
This is consistent with Fig. 3 in \cite{kasuya-2021} in relation to the study of the shadow of the nonspherical symmetric wormholes.

\section{Summary and outlook}
\label{secVI}
In this paper, we study the strong gravitational lensing by Kerr-like wormholes with an additional parameter $\lambda$ compared to Kerr black holes, which is used to specify the location of the throat.
The corresponding radial potential is derived from the null geodesics with six roots. 
Four of them depend on the mass and the spin of the wormholes as well as the impact parameter of the light rays, and the other two roots are determined by the throat and the root in (\ref{r_root}).
We classify the roots and construct the parameter space diagram in terms of the impact parameter of the light rays and the parameter $\lambda$ of the throat, where the throat, together with other roots, becomes either a double or triple root, potentially giving a divergence of the deflection angle in the SDL.
In addition, the effective potential as a function of the proper distance from the throat is constructed, with which one can see how the light rays can either travel within a single spacetime, where both sources and observers are located, or pass from the source through the throat into another spacetime where different observers reside.
When the observers and the sources are in different spacetimes, the deflection angle can still be defined using the symmetry of the proper distance $l \to -l$ that maps the light rays from the spacetime of the sources to the spacetime of the observers.
It is assumed that the same gravitational lensing equation can be applied as in the cases where both observers and the light sources are in the same spacetime.
In general, while the logarithmic divergence is known to occur as the closest distance of the incident light rays around a black hole $r_0$ approaches a double root, the stronger power-law (nonlogarithmic) divergence is found as $r_0$ approaches the triple root, especially in a wormhole.
%
%
In the future, the most prompt work for us is by following \cite{taylor-2014} to find the trajectories of light rays, perhaps analytically.

\appendix
\section{Roots of the radial potential}
\label{app}
It is of importance to realize the properties of the roots of the radial potential $\tilde R(r)$, with which to construct the diagram of the parameter space in Fig. \ref{parameter_space} in terms of the impact parameter $b$ of light rays and the parameter $\lambda$ associated with the wormholes.
As shown in (\ref{R_tilde}), the roots of the radial potential $\tilde R(r)$ can come from its counterpart $R(r)$ in the case of Kerr black holes.
It will be very helpful to summarize them according to \cite{wang-2022}.
The radial potential $R(r)$ of the Kerr black holes on the equatorial plane can be rewritten as a quartic function
\begin{align}
R(r)=r^4+Ur^2+Vr
\end{align}
with the coefficient functions given by
\begin{align}
&U=a^2-b^2\;,\\
&V=2M\left(b_s -a\right)^2\;.
\end{align}
There are four roots,
namely $R(r)=(r-r_1)(r-r_2) (r-r_3) (r-r_4)$ with the properties $r_{1}+r_{2}+r_{3}+r_{4}=0$ and $r_1 <r_2 < r_3\le r_4$, and can be written as
\begin{align}
r_{1}&=-z-\sqrt{-\hspace*{1mm}\frac{U}{2}-z^2+\frac{V}{4z}}\;,\\
r_{2}&=-z+\sqrt{-\hspace*{1mm}\frac{U}{2}-z^2+\frac{V}{4z}}\;,\\
r_{3}&=+z-\sqrt{-\hspace*{1mm}\frac{U}{2}-z^2-\frac{V}{4z}}\;,\label{r_3}\\
r_{4}&=+z+\sqrt{-\hspace*{1mm}\frac{U}{2}-z^2-\frac{V}{4z}}\;.\label{r_4}
\end{align}
The following notation has been used,
\begin{align}
z&=\sqrt{\frac{\Omega_{+}+\Omega_{-}-\frac{U}{3}}{2}}\, , \quad \quad \Omega_{\pm}=\sqrt[3]{-\hspace*{1mm}\frac{\varkappa}{2}\pm\sqrt{\left(\frac{\varpi}{3}\right)^3+\left(\frac{\varkappa}{2}\right)^2}} \;,
\end{align}
where
\begin{align}
\mathcal{\varpi}=-\hspace*{1mm}\frac{U^2}{12}-W \, , \quad\quad
\mathcal{\varkappa}=-\hspace*{1mm}\frac{U}{3}\left[\left(\frac{U}{6}\right)^2-W\right]-\hspace*{1mm}\frac{V^2}{8}\,.
\end{align}
Apparently, using the above formulas gives $r_2=0$ as anticipated.

Given the parameters $b$ and $\lambda$, we can determine whether the roots are real or complex, and also the corresponding turning point of the light rays to categorize various types of orbits.

\section{General Kerr spacetime and one parameter Kerr-like wormhole}
\label{secV_add}
It is known that the Kerr solution is not general enough to describe the exterior spacetime geometry of all astrophysical rotating compact objects.
Therefore, the generalization of the Kerr spacetime is useful for fitting future observational data.
In \cite{visser-2023}, a more general distortion of Kerr spacetime was proposed by imposing the conditions of Hamilton-Jacobi timelike separability, Klein-Gordon separability, and asymptotic flatness (for more details, see \cite{visser-2023}).
Let us now consider the 3-function generalization of the Kerr spacetime in \cite{visser-2023} with the line element written as
\begin{equation}
\begin{split}\label{2_fun}
ds^2= & -\frac{\Delta_V(r)\exp(-2\Phi_V(r))-a^2\sin^2\theta}{\Xi_V(r)^2+a^2\cos^2\theta}dt^2
+\frac{\Xi_V(r)^2+a^2\cos^2\theta}{\Delta_V(r)}dr^2\\
&+\left[ \Xi_V(r)^2+a^2\cos^2\theta \right] d\theta^2
+\frac{\left[ (\Xi_V(r)^2+a^2)^2-\exp(-2\Phi_V(r))\Delta_V(r)a^2\sin^2\theta \right]\sin^2\theta}{\Xi_V(r)^2+a^2\cos^2\theta}d\phi^2\\
&-\frac{2a \left[ \Xi_V(r)^2-\Delta_V(r)\exp(-2\Phi_V(r))+a^2 \right] \sin^2\theta}{\Xi_V(r)^2+a^2\cos^2\theta}dtd\phi ,
\end{split}
\end{equation}
where $\Xi_V(r)$, $\Delta_V(r)$, and $\Phi_V(r)$, three functions of the radial coordinate $r$, are introduced.
The particular type of the Kerr-like wormhole metric (\ref{KN_metric}) in this paper can be reduced from (\ref{2_fun}) with the following identifications
\be
\Xi_V(r) = r \, {\rm ,}\quad\quad
\Delta_V(r) = \hat{\Delta}(r) \, {\rm ,}\quad\quad
e^{-2\Phi_V(r)} = \frac{\Delta(r)}{\hat{\Delta}(r)}
\ee
in terms of one parameter $\lambda$ used to specify the location of the throat.
In the case of $\hat{\Delta}(r)=\Delta(r)$ for $\lambda=0$, the metric reduces to the Kerr metric.
The light deflection effects can then be compared between Kerr black holes and Kerr-like wormholes.
There are some other types of wormholes, which can be described by the 3-function generalization by choosing the appropriate functions, as shown in \cite{visser-2023}.
It will be extremely interesting to consider the deflection of light by the astrophysical rotating objects in the more general metric of (\ref{2_fun}).
Given the metric (\ref{2_fun}), the equations of motion for the light on the equatorial plane with the impact parameter $b_s$ and the energy $\varepsilon$ now become
\begin{align}
&\frac{\Xi_V(r)^2}{\varepsilon}\frac{dr}{d\sigma}=\pm_r\sqrt{\frac{\Xi_V(r)^4 +2a(a-b_s)\Xi_V(r)^2 +(a-b_s)^2 [a^2-\Delta_V(r) e^{-2\Phi_V(r)}]}{\exp(-2\Phi_V(r))} }\, ,\\
&\frac{\Xi_V(r)^2}{\varepsilon}\frac{d\phi}{d\sigma} =\frac{a}{\Delta_V(r) \exp(-2\Phi_V(r))} \left[\Xi_V(r)^2 +a^2 -a b_s\right] +b_s -a \, ,\\
&\frac{\Xi_V(r)^2}{\varepsilon}\frac{dt}{d\sigma}=\frac{\Xi_V(r)^2+a^2}{\Delta_V(r) \exp(-2\Phi_V(r))} \left[\Xi_V(r)^2+a^2-a b_s\right] + a\left(b_s-a\right) \,
\end{align}
from which, one can derive the corresponding equation as in (\ref{dr/dphi}), integrate it along the coordinate $r$ from the sources, through the turning point, and up to the observers, to find the deflection angle.

\begin{acknowledgments}
This work was supported in part by the National Science and Technology Council (NSTC) of Taiwan, Republic of China.
\end{acknowledgments}

\addcontentsline{toc}{chapter}{Bibliography}
\bibliography{References}

\begin{thebibliography}{44}
\expandafter\ifx\csname natexlab\endcsname\relax\def\natexlab#1{#1}\fi
\expandafter\ifx\csname bibnamefont\endcsname\relax
  \def\bibnamefont#1{#1}\fi
\expandafter\ifx\csname bibfnamefont\endcsname\relax
  \def\bibfnamefont#1{#1}\fi
\expandafter\ifx\csname citenamefont\endcsname\relax
  \def\citenamefont#1{#1}\fi
\expandafter\ifx\csname url\endcsname\relax
  \def\url#1{\texttt{#1}}\fi
\expandafter\ifx\csname urlprefix\endcsname\relax\def\urlprefix{URL }\fi
\providecommand{\bibinfo}[2]{#2}
\providecommand{\eprint}[2][]{\url{#2}}

\bibitem[{\citenamefont{Misner et~al.}(1974)\citenamefont{Misner, Thorne, Wheeler, and Chandrasekhar}}]{misner-1974}
\bibinfo{author}{\bibfnamefont{C.~W.} \bibnamefont{Misner}}, \bibinfo{author}{\bibfnamefont{K.~S.} \bibnamefont{Thorne}}, \bibinfo{author}{\bibfnamefont{J.~A.} \bibnamefont{Wheeler}}, \bibnamefont{and} \bibinfo{author}{\bibfnamefont{S.}~\bibnamefont{Chandrasekhar}}, \bibinfo{journal}{Phys. Today} \textbf{\bibinfo{volume}{27}}, \bibinfo{pages}{No. 8, 47} (\bibinfo{year}{1974}), \urlprefix\url{https://doi.org/10.1063/1.3128805}.

\bibitem[{\citenamefont{Einstein and Rosen}(1935)}]{einstein-1935}
\bibinfo{author}{\bibfnamefont{A.}~\bibnamefont{Einstein}} \bibnamefont{and} \bibinfo{author}{\bibfnamefont{N.}~\bibnamefont{Rosen}}, \bibinfo{journal}{Phys. Rev.} \textbf{\bibinfo{volume}{48}}, \bibinfo{pages}{73} (\bibinfo{year}{1935}), \urlprefix\url{https://doi.org/10.1103/physrev.48.73}.

\bibitem[{\citenamefont{Wheeler}(1955)}]{wheeler-1955}
\bibinfo{author}{\bibfnamefont{J.~A.} \bibnamefont{Wheeler}}, \bibinfo{journal}{Phys. Rev.} \textbf{\bibinfo{volume}{97}}, \bibinfo{pages}{511} (\bibinfo{year}{1955}), \urlprefix\url{https://doi.org/10.1103/physrev.97.511}.

\bibitem[{\citenamefont{Wheeler}(1962)}]{wheeler-1962}
\bibinfo{author}{\bibfnamefont{J.~A.} \bibnamefont{Wheeler}}, \emph{\bibinfo{title}{{Geometrodynamics}}} (\bibinfo{publisher}{Academic Press, Boston}, \bibinfo{year}{1962}).

\bibitem[{\citenamefont{Morris et~al.}(1988)\citenamefont{Morris, Thorne, and Yurtsever}}]{morris-1988}
\bibinfo{author}{\bibfnamefont{M.~S.} \bibnamefont{Morris}}, \bibinfo{author}{\bibfnamefont{K.~S.} \bibnamefont{Thorne}}, \bibnamefont{and} \bibinfo{author}{\bibfnamefont{U.}~\bibnamefont{Yurtsever}}, \bibinfo{journal}{Phys. Rev. Lett} \textbf{\bibinfo{volume}{61}}, \bibinfo{pages}{1446} (\bibinfo{year}{1988}), \urlprefix\url{https://doi.org/10.1103/physrevlett.61.1446}.

\bibitem[{\citenamefont{Visser}(1995)}]{Visser-1995}
\bibinfo{author}{\bibfnamefont{M.}~\bibnamefont{Visser}}, \emph{\bibinfo{title}{{Lorentzian Wormholes}}} (\bibinfo{publisher}{American Institute of Physics, New York}, \bibinfo{year}{1995}).

\bibitem[{\citenamefont{\text{Abbott}~\textit{et al.} \text{(LIGO Scientific and Virgo Collaboration)}}(2016)}]{abbott-2016}
\bibinfo{author}{\bibfnamefont{B.~P.} \bibnamefont{\text{Abbott}~\textit{et al.} \text{(LIGO Scientific and Virgo Collaboration)}}}, \bibinfo{journal}{Phys. Rev. Lett} \textbf{\bibinfo{volume}{116}}, \bibinfo{pages}{061102} (\bibinfo{year}{2016}), \urlprefix\url{https://doi.org/10.1103/physrevlett.116.061102}.

\bibitem[{\citenamefont{\text{Abbott}~\textit{et al.} \text{(LIGO Scientific and Virgo Collaboration)}}(2019)}]{abbott-2019}
\bibinfo{author}{\bibfnamefont{B.~P.} \bibnamefont{\text{Abbott}~\textit{et al.} \text{(LIGO Scientific and Virgo Collaboration)}}}, \bibinfo{journal}{Phys. Rev. X} \textbf{\bibinfo{volume}{9}}, \bibinfo{pages}{031040} (\bibinfo{year}{2019}), \urlprefix\url{https://doi.org/10.1103/physrevx.9.031040}.

\bibitem[{\citenamefont{\text{Abbott}~\textit{et al.} \text{(LIGO Scientific and Virgo Collaboration)}}(2021)}]{abbott-2021}
\bibinfo{author}{\bibfnamefont{R.}~\bibnamefont{\text{Abbott}~\textit{et al.} \text{(LIGO Scientific and Virgo Collaboration)}}}, \bibinfo{journal}{Phys. Rev. X} \textbf{\bibinfo{volume}{11}}, \bibinfo{pages}{021053} (\bibinfo{year}{2021}), \urlprefix\url{https://doi.org/10.1103/physrevx.11.021053}.

\bibitem[{\citenamefont{\text{The Event Horizon Telescope Collaboration}}(2019)}]{akiyama-2019}
\bibinfo{author}{\bibnamefont{\text{The Event Horizon Telescope Collaboration}}}, \bibinfo{journal}{Astrophys. J. Lett.} \textbf{\bibinfo{volume}{875}}, \bibinfo{pages}{L1} (\bibinfo{year}{2019}), \urlprefix\url{https://doi.org/10.3847/2041-8213/ab0ec7}.

\bibitem[{\citenamefont{\text{The Event Horizon Telescope Collaboration}}(2022)}]{collaboration-2022}
\bibinfo{author}{\bibnamefont{\text{The Event Horizon Telescope Collaboration}}}, \bibinfo{journal}{Astrophys. J. Lett.} \textbf{\bibinfo{volume}{930}}, \bibinfo{pages}{L12} (\bibinfo{year}{2022}), \urlprefix\url{https://doi.org/10.3847/2041-8213/ac6674}.

\bibitem[{\citenamefont{Hartle}(2003)}]{hartle-2003}
\bibinfo{author}{\bibfnamefont{J.~B.} \bibnamefont{Hartle}}, \emph{\bibinfo{title}{Gravity: An Introduction to Einstein's General Relativity}} (\bibinfo{publisher}{Benjamin Cummings, New York}, \bibinfo{year}{2003}).

\bibitem[{\citenamefont{Virbhadra and Ellis}(2000)}]{virbhadra-2000}
\bibinfo{author}{\bibfnamefont{K.~S.} \bibnamefont{Virbhadra}} \bibnamefont{and} \bibinfo{author}{\bibfnamefont{G.~F.~R.} \bibnamefont{Ellis}}, \bibinfo{journal}{Phys. Rev. D} \textbf{\bibinfo{volume}{62}}, \bibinfo{pages}{084003} (\bibinfo{year}{2000}), \urlprefix\url{https://doi.org/10.1103/physrevd.62.084003}.

\bibitem[{\citenamefont{Frittelli et~al.}(2000)\citenamefont{Frittelli, Kling, and Newman}}]{frittelli-2000}
\bibinfo{author}{\bibfnamefont{S.}~\bibnamefont{Frittelli}}, \bibinfo{author}{\bibfnamefont{T.~P.} \bibnamefont{Kling}}, \bibnamefont{and} \bibinfo{author}{\bibfnamefont{E.~T.} \bibnamefont{Newman}}, \bibinfo{journal}{Phys. Rev. D} \textbf{\bibinfo{volume}{61}}, \bibinfo{pages}{064021} (\bibinfo{year}{2000}), \urlprefix\url{https://doi.org/10.1103/physrevd.61.064021}.

\bibitem[{\citenamefont{Bozza et~al.}(2001)\citenamefont{Bozza, Capozziello, Iovane, and Scarpetta}}]{bozza-2001}
\bibinfo{author}{\bibfnamefont{V.}~\bibnamefont{Bozza}}, \bibinfo{author}{\bibfnamefont{S.}~\bibnamefont{Capozziello}}, \bibinfo{author}{\bibfnamefont{G.}~\bibnamefont{Iovane}}, \bibnamefont{and} \bibinfo{author}{\bibfnamefont{G.}~\bibnamefont{Scarpetta}}, \bibinfo{journal}{Gen. Relativ. Gravit.} \textbf{\bibinfo{volume}{33}}, \bibinfo{pages}{1535} (\bibinfo{year}{2001}), \urlprefix\url{https://doi.org/10.1023/a:1012292927358}.

\bibitem[{\citenamefont{Bozza}(2002)}]{bozza-2002}
\bibinfo{author}{\bibfnamefont{V.}~\bibnamefont{Bozza}}, \bibinfo{journal}{Phys. Rev. D} \textbf{\bibinfo{volume}{66}}, \bibinfo{pages}{103001} (\bibinfo{year}{2002}), \urlprefix\url{https://doi.org/10.1103/physrevd.66.103001}.

\bibitem[{\citenamefont{Eiroa et~al.}(2002)\citenamefont{Eiroa, Romero, and Torres}}]{eiroa-2002}
\bibinfo{author}{\bibfnamefont{E.~F.} \bibnamefont{Eiroa}}, \bibinfo{author}{\bibfnamefont{G.~E.} \bibnamefont{Romero}}, \bibnamefont{and} \bibinfo{author}{\bibfnamefont{D.~F.} \bibnamefont{Torres}}, \bibinfo{journal}{Phys. Rev. D} \textbf{\bibinfo{volume}{66}}, \bibinfo{pages}{024010} (\bibinfo{year}{2002}), \urlprefix\url{https://doi.org/10.1103/physrevd.66.024010}.

\bibitem[{\citenamefont{Iyer and Petters}(2007)}]{iyer-2007}
\bibinfo{author}{\bibfnamefont{S.~V.} \bibnamefont{Iyer}} \bibnamefont{and} \bibinfo{author}{\bibfnamefont{A.~O.} \bibnamefont{Petters}}, \bibinfo{journal}{Gen. Relativ. Gravit.} \textbf{\bibinfo{volume}{39}}, \bibinfo{pages}{1563} (\bibinfo{year}{2007}), \urlprefix\url{https://doi.org/10.1007/s10714-007-0481-8}.

\bibitem[{\citenamefont{Iyer and Hansen}(2009)}]{iyer-2009}
\bibinfo{author}{\bibfnamefont{S.~V.} \bibnamefont{Iyer}} \bibnamefont{and} \bibinfo{author}{\bibfnamefont{E.~C.} \bibnamefont{Hansen}}, \bibinfo{journal}{arXiv: 0908.0085}  (\bibinfo{year}{2009}), \urlprefix\url{https://arxiv.org/abs/0908.0085}.

\bibitem[{\citenamefont{Tsukamoto}(2017)}]{tsukamoto-2017}
\bibinfo{author}{\bibfnamefont{N.}~\bibnamefont{Tsukamoto}}, \bibinfo{journal}{Phys. Rev. D} \textbf{\bibinfo{volume}{95}}, \bibinfo{pages}{064035} (\bibinfo{year}{2017}), \urlprefix\url{https://doi.org/10.1103/physrevd.95.064035}.

\bibitem[{\citenamefont{Hsiao et~al.}(2020)\citenamefont{Hsiao, Lee, and Lin}}]{hsiao-2020}
\bibinfo{author}{\bibfnamefont{Y.-W.} \bibnamefont{Hsiao}}, \bibinfo{author}{\bibfnamefont{D.-S.} \bibnamefont{Lee}}, \bibnamefont{and} \bibinfo{author}{\bibfnamefont{C.-Y.} \bibnamefont{Lin}}, \bibinfo{journal}{Phys. Rev. D} \textbf{\bibinfo{volume}{101}}, \bibinfo{pages}{064070} (\bibinfo{year}{2020}), \urlprefix\url{https://doi.org/10.1103/physrevd.101.064070}.

\bibitem[{\citenamefont{Hsieh et~al.}(2021{\natexlab{a}})\citenamefont{Hsieh, Lee, and Lin}}]{hsieh-2021A}
\bibinfo{author}{\bibfnamefont{T.}~\bibnamefont{Hsieh}}, \bibinfo{author}{\bibfnamefont{D.-S.} \bibnamefont{Lee}}, \bibnamefont{and} \bibinfo{author}{\bibfnamefont{C.-Y.} \bibnamefont{Lin}}, \bibinfo{journal}{Phys. Rev. D} \textbf{\bibinfo{volume}{103}}, \bibinfo{pages}{104063} (\bibinfo{year}{2021}{\natexlab{a}}), \urlprefix\url{https://doi.org/10.1103/physrevd.103.104063}.

\bibitem[{\citenamefont{Hsieh et~al.}(2021{\natexlab{b}})\citenamefont{Hsieh, Lee, and Lin}}]{hsieh-2021B}
\bibinfo{author}{\bibfnamefont{T.}~\bibnamefont{Hsieh}}, \bibinfo{author}{\bibfnamefont{D.-S.} \bibnamefont{Lee}}, \bibnamefont{and} \bibinfo{author}{\bibfnamefont{C.-Y.} \bibnamefont{Lin}}, \bibinfo{journal}{Phys. Rev. D} \textbf{\bibinfo{volume}{104}}, \bibinfo{pages}{104013} (\bibinfo{year}{2021}{\natexlab{b}}), \urlprefix\url{https://doi.org/10.1103/physrevd.104.104013}.

\bibitem[{\citenamefont{Damour and Solodukhin}(2007)}]{damour-2007}
\bibinfo{author}{\bibfnamefont{T.}~\bibnamefont{Damour}} \bibnamefont{and} \bibinfo{author}{\bibfnamefont{S.~N.} \bibnamefont{Solodukhin}}, \bibinfo{journal}{Phys. Rev. D} \textbf{\bibinfo{volume}{76}}, \bibinfo{pages}{024016} (\bibinfo{year}{2007}), \urlprefix\url{https://doi.org/10.1103/physrevd.76.024016}.

\bibitem[{\citenamefont{Tsukamoto}(2020)}]{tsukamoto-2020}
\bibinfo{author}{\bibfnamefont{N.}~\bibnamefont{Tsukamoto}}, \bibinfo{journal}{Phys. Rev. D} \textbf{\bibinfo{volume}{101}}, \bibinfo{pages}{104021} (\bibinfo{year}{2020}), \urlprefix\url{https://doi.org/10.1103/physrevd.101.104021}.

\bibitem[{\citenamefont{Övgün}(2018)}]{ovgun-2018}
\bibinfo{author}{\bibfnamefont{A.}~\bibnamefont{Övgün}}, \bibinfo{journal}{Phys. Rev. D} \textbf{\bibinfo{volume}{98}}, \bibinfo{pages}{044033} (\bibinfo{year}{2018}), \urlprefix\url{https://doi.org/10.1103/physrevd.98.044033}.

\bibitem[{\citenamefont{Kasuya and Kobayashi}(2021)}]{kasuya-2021}
\bibinfo{author}{\bibfnamefont{S.}~\bibnamefont{Kasuya}} \bibnamefont{and} \bibinfo{author}{\bibfnamefont{M.}~\bibnamefont{Kobayashi}}, \bibinfo{journal}{Phys. Rev. D} \textbf{\bibinfo{volume}{103}}, \bibinfo{pages}{104050} (\bibinfo{year}{2021}), \urlprefix\url{https://doi.org/10.1103/physrevd.103.104050}.

\bibitem[{\citenamefont{Bueno et~al.}(2018)\citenamefont{Bueno, Cano, Goelen, Hertog, and Vercnocke}}]{bueno-2018}
\bibinfo{author}{\bibfnamefont{P.}~\bibnamefont{Bueno}}, \bibinfo{author}{\bibfnamefont{P.~A.} \bibnamefont{Cano}}, \bibinfo{author}{\bibfnamefont{F.}~\bibnamefont{Goelen}}, \bibinfo{author}{\bibfnamefont{T.}~\bibnamefont{Hertog}}, \bibnamefont{and} \bibinfo{author}{\bibfnamefont{B.}~\bibnamefont{Vercnocke}}, \bibinfo{journal}{Phys. Rev. D} \textbf{\bibinfo{volume}{97}}, \bibinfo{pages}{024040} (\bibinfo{year}{2018}), \urlprefix\url{https://doi.org/10.1103/physrevd.97.024040}.

\bibitem[{\citenamefont{Amir et~al.}(2019)\citenamefont{Amir, Jusufi, Banerjee, and Hansraj}}]{amir-2019}
\bibinfo{author}{\bibfnamefont{M.}~\bibnamefont{Amir}}, \bibinfo{author}{\bibfnamefont{K.}~\bibnamefont{Jusufi}}, \bibinfo{author}{\bibfnamefont{A.}~\bibnamefont{Banerjee}}, \bibnamefont{and} \bibinfo{author}{\bibfnamefont{S.}~\bibnamefont{Hansraj}}, \bibinfo{journal}{Classical Quantum Gravity} \textbf{\bibinfo{volume}{36}}, \bibinfo{pages}{215007} (\bibinfo{year}{2019}), \urlprefix\url{https://doi.org/10.1088/1361-6382/ab42be}.

\bibitem[{\citenamefont{Carter}(1968)}]{carter-1968}
\bibinfo{author}{\bibfnamefont{B.}~\bibnamefont{Carter}}, \bibinfo{journal}{Commun. Math. Phys.} \textbf{\bibinfo{volume}{10}}, \bibinfo{pages}{280} (\bibinfo{year}{1968}), \urlprefix\url{https://doi.org/10.1007/bf03399503}.

\bibitem[{\citenamefont{Frolov et~al.}(2007)\citenamefont{Frolov, Krtouš, and Kubiznák}}]{frolov-2007}
\bibinfo{author}{\bibfnamefont{V.~P.} \bibnamefont{Frolov}}, \bibinfo{author}{\bibfnamefont{P.}~\bibnamefont{Krtouš}}, \bibnamefont{and} \bibinfo{author}{\bibfnamefont{D.}~\bibnamefont{Kubiznák}}, \bibinfo{journal}{J. High Energy Phys.} \textbf{\bibinfo{volume}{02}}, \bibinfo{pages}{005} (\bibinfo{year}{2007}), \urlprefix\url{https://doi.org/10.1088/1126-6708/2007/02/005}.

\bibitem[{\citenamefont{Frolov et~al.}(2017)\citenamefont{Frolov, Krtouš, and Kubizňák}}]{frolov-2017}
\bibinfo{author}{\bibfnamefont{V.~P.} \bibnamefont{Frolov}}, \bibinfo{author}{\bibfnamefont{P.}~\bibnamefont{Krtouš}}, \bibnamefont{and} \bibinfo{author}{\bibfnamefont{D.}~\bibnamefont{Kubizňák}}, \bibinfo{journal}{Living Rev. Relativity} \textbf{\bibinfo{volume}{20}}, \bibinfo{pages}{6} (\bibinfo{year}{2017}), \urlprefix\url{https://doi.org/10.1007/s41114-017-0009-9}.

\bibitem[{\citenamefont{Baines and Visser}(2023)}]{visser-2023}
\bibinfo{author}{\bibfnamefont{J.}~\bibnamefont{Baines}} \bibnamefont{and} \bibinfo{author}{\bibfnamefont{M.}~\bibnamefont{Visser}}, \bibinfo{journal}{Universe} \textbf{\bibinfo{volume}{9}}, \bibinfo{pages}{223} (\bibinfo{year}{2023}), \urlprefix\url{https://doi.org/10.3390/universe9050223}.

\bibitem[{\citenamefont{Taylor}(2014)}]{taylor-2014}
\bibinfo{author}{\bibfnamefont{P.}~\bibnamefont{Taylor}}, \bibinfo{journal}{Phys. Rev. D} \textbf{\bibinfo{volume}{90}}, \bibinfo{pages}{024057} (\bibinfo{year}{2014}), \urlprefix\url{https://doi.org/10.1103/physrevd.90.024057}.

\bibitem[{\citenamefont{Shaikh et~al.}(2019)\citenamefont{Shaikh, Banerjee, Paul, and Sarkar}}]{shaikh-2019}
\bibinfo{author}{\bibfnamefont{R.}~\bibnamefont{Shaikh}}, \bibinfo{author}{\bibfnamefont{P.}~\bibnamefont{Banerjee}}, \bibinfo{author}{\bibfnamefont{S.}~\bibnamefont{Paul}}, \bibnamefont{and} \bibinfo{author}{\bibfnamefont{T.}~\bibnamefont{Sarkar}}, \bibinfo{journal}{J. Cosmol. Astropart. Phys.} \textbf{\bibinfo{volume}{07}}, \bibinfo{pages}{028} (\bibinfo{year}{2019}), \urlprefix\url{https://doi.org/10.1088/1475-7516/2019/07/028}.

\bibitem[{\citenamefont{Tsukamoto}(2023)}]{tsukamoto-2023}
\bibinfo{author}{\bibfnamefont{N.}~\bibnamefont{Tsukamoto}}, \bibinfo{journal}{Classical Quantum Gravity} \textbf{\bibinfo{volume}{40}}, \bibinfo{pages}{228001} (\bibinfo{year}{2023}), \urlprefix\url{https://doi.org/10.1088/1361-6382/acfb6e}.

\bibitem[{\citenamefont{\text{Vagnozzi} \textit{et al.}}(2023)}]{vagnozzi-2023}
\bibinfo{author}{\bibfnamefont{S.}~\bibnamefont{\text{Vagnozzi} \textit{et al.}}}, \bibinfo{journal}{Classical Quantum Gravity} \textbf{\bibinfo{volume}{40}}, \bibinfo{pages}{165007} (\bibinfo{year}{2023}), \urlprefix\url{https://doi.org/10.1088/1361-6382/acd97b}.

\bibitem[{\citenamefont{Dai and Stojkovic}(2019)}]{dai-2019}
\bibinfo{author}{\bibfnamefont{D.-C.} \bibnamefont{Dai}} \bibnamefont{and} \bibinfo{author}{\bibfnamefont{D.}~\bibnamefont{Stojkovic}}, \bibinfo{journal}{Phys. Rev. D} \textbf{\bibinfo{volume}{100}}, \bibinfo{pages}{083513} (\bibinfo{year}{2019}), \urlprefix\url{https://doi.org/10.1103/physrevd.100.083513}.

\bibitem[{\citenamefont{Simonetti et~al.}(2021)\citenamefont{Simonetti, Kavic, Minic, Stojkovic, and Dai}}]{simonetti-2021}
\bibinfo{author}{\bibfnamefont{J.~H.} \bibnamefont{Simonetti}}, \bibinfo{author}{\bibfnamefont{M.~J.} \bibnamefont{Kavic}}, \bibinfo{author}{\bibfnamefont{D.}~\bibnamefont{Minic}}, \bibinfo{author}{\bibfnamefont{D.}~\bibnamefont{Stojkovic}}, \bibnamefont{and} \bibinfo{author}{\bibfnamefont{D.-C.} \bibnamefont{Dai}}, \bibinfo{journal}{Phys. Rev. D} \textbf{\bibinfo{volume}{104}}, \bibinfo{pages}{L081502} (\bibinfo{year}{2021}), \urlprefix\url{https://doi.org/10.1103/physrevd.104.l081502}.

\bibitem[{\citenamefont{Bambi and Stojkovic}(2021)}]{bambi-2021}
\bibinfo{author}{\bibfnamefont{C.}~\bibnamefont{Bambi}} \bibnamefont{and} \bibinfo{author}{\bibfnamefont{D.}~\bibnamefont{Stojkovic}}, \bibinfo{journal}{Universe} \textbf{\bibinfo{volume}{7}}, \bibinfo{pages}{136} (\bibinfo{year}{2021}), \urlprefix\url{https://doi.org/10.3390/universe7050136}.

\bibitem[{\citenamefont{Gralla et~al.}(2018)\citenamefont{Gralla, Lupsasca, and Strominger}}]{gralla-2018}
\bibinfo{author}{\bibfnamefont{S.~E.} \bibnamefont{Gralla}}, \bibinfo{author}{\bibfnamefont{A.}~\bibnamefont{Lupsasca}}, \bibnamefont{and} \bibinfo{author}{\bibfnamefont{A.}~\bibnamefont{Strominger}}, \bibinfo{journal}{Mon. Not. R. Astron. Soc.} \textbf{\bibinfo{volume}{475}}, \bibinfo{pages}{3829} (\bibinfo{year}{2018}), \urlprefix\url{https://doi.org/10.1093/mnras/sty039}.

\bibitem[{\citenamefont{Gralla and Lupsasca}(2020)}]{gralla-2020}
\bibinfo{author}{\bibfnamefont{S.~E.} \bibnamefont{Gralla}} \bibnamefont{and} \bibinfo{author}{\bibfnamefont{A.}~\bibnamefont{Lupsasca}}, \bibinfo{journal}{Phys. Rev. D} \textbf{\bibinfo{volume}{101}}, \bibinfo{pages}{044032} (\bibinfo{year}{2020}), \urlprefix\url{https://doi.org/10.1103/physrevd.101.044032}.

\bibitem[{\citenamefont{Wang et~al.}(2022)\citenamefont{Wang, Lee, and Lin}}]{wang-2022}
\bibinfo{author}{\bibfnamefont{C.-Y.} \bibnamefont{Wang}}, \bibinfo{author}{\bibfnamefont{D.-S.} \bibnamefont{Lee}}, \bibnamefont{and} \bibinfo{author}{\bibfnamefont{C.-Y.} \bibnamefont{Lin}}, \bibinfo{journal}{Phys. Rev. D} \textbf{\bibinfo{volume}{106}}, \bibinfo{pages}{084048} (\bibinfo{year}{2022}), \urlprefix\url{https://doi.org/10.1103/physrevd.106.084048}.

\bibitem[{\citenamefont{Chandrasekhar}(1992)}]{chandrasekhar-1998}
\bibinfo{author}{\bibfnamefont{S.}~\bibnamefont{Chandrasekhar}}, \emph{\bibinfo{title}{{The Mathematical Theory of Black Holes}}} (\bibinfo{publisher}{Oxford University Press, New York}, \bibinfo{year}{1992}).

\end{thebibliography}

\end{document}